\documentclass[journal,10pt,twocolumn,twoside]{IEEEtran}
\usepackage{multicol}   
\usepackage{multirow}
\usepackage[pdftex]{graphicx}
\usepackage{algorithm}
\usepackage{algpseudocode}
\usepackage[small]{caption}
\usepackage{float}
\usepackage{amssymb,amsmath}
\usepackage{graphicx}
\usepackage{subfigure}
\usepackage{xspace}
\usepackage{amsthm}
\usepackage{fancyhdr}

\usepackage{caption,setspace}
\usepackage{amssymb,amsmath}
\usepackage{bm}
\usepackage{textcomp} 
\usepackage{epstopdf}
\usepackage{bbding}
\usepackage{cite} 
\usepackage{color}
\usepackage{pifont}
\usepackage{lipsum}
\usepackage{amstext}
\usepackage{orcidlink}

\usepackage{cuted}
\usepackage{stfloats}

\graphicspath{{fig/}}

\newcommand\scalemath[2]{\scalebox{#1}{\mbox{\ensuremath{\displaystyle #2}}}}

\begin{document}

\title{\huge Leveraging Movable-Element STARS for RSMA-SWIPT Under Practical System Imperfections}
\author{Muhammad Asif \,\orcidlink{0000-0002-9699-1675}, \IEEEmembership{Member, IEEE}, Asim Ihsan \,\orcidlink{0000-0001-7491-7178}, Irfan Muhammad \,\orcidlink{0000-0001-5703-7988}, \IEEEmembership{Member, IEEE}, Zhu Shoujin \,\orcidlink{0000-0002-2615-2489}, Mikael Gidlund \,\orcidlink{0000-0003-0873-7827}, \IEEEmembership{Fellow, IEEE}, and Symeon Chatzinotas \,\orcidlink{0000-0001-5122-0001}, \IEEEmembership{Fellow, IEEE} 

\thanks{This work was supported in part by the Tongling University Talent Research Initiation Fund Project under Grant No. 2023tlxyrc17; and in part by the Anhui Provincial Higher Education Institutions Young Core Faculty Domestic Visiting Scholar Training Funding Program under Grant No. JNFX2025067.}  

\thanks{Muhammad Asif and Zhu Shoujin are with the School of Electrical and Information Engineering, Tongling University, Tongling 244002, China (e-mails: masif@tlu.edu.cn, 2023028@tlu.edu.cn).
	
 Asim Ihsan is with the Interdisciplinary Research Center for Communication Systems and Sensing, King Fahd University of Petroleum \& Minerals (KFUPM ), Dhahran, Saudi Arabia (e-mail: asim.ihsan@kfupm.edu.sa).
	
 Irfan Muhammad is with the Centre for Wireless Communications, University of Oulu, 90570 Oulu, Finland (e-mail: irfan.muhammad@oulu.fi).
  
M. Gidlund is with the Department of Computer and Electrical Engineering, Mid Sweden University, Homlgatan 10, 851 70 Sundsvall, Sweden (e-mail: mikael.gidlund@miun.se).
  	
 Symeon Chatzinotas is with the Interdisciplinary Centre for Security, Reliability and Trust (SnT), University of Luxembourg, 1855 Luxembourg City, Luxembourg (e-mail: symeon.chatzinotas@uni.lu).

}

\vspace{-0.7cm}}%

\markboth{IEEE TRANSACTIONS ON VEHICULAR TECHNOLOGY}
{ \MakeLowercase{\textit{}}} 
\maketitle

\begin{abstract}
We study a robust multiuser framework for simultaneous wireless information and power transfer (SWIPT), where rate-splitting multiple access (RSMA) is integrated with a movable-element simultaneously transmitting and reflecting surface (ME-STARS). By allowing the STARS elements to change their positions in addition to controlling the reflection and transmission coefficients, the proposed architecture provides additional spatial flexibility for improving the cascaded channels toward users in both regions. Meanwhile, RSMA is employed to manage multiuser interference, while power-splitting receivers enable simultaneous information decoding and energy harvesting. The system design further accounts for channel state information (CSI) uncertainty, residual transceiver hardware impairments (HIs), and the nonlinear characteristics of practical energy-harvesting circuits. Accordingly, a robust sum-rate maximization problem is formulated by jointly designing the base-station precoders, common-rate allocation, ME-STARS reflection/transmission coefficients, power-splitting (PS) ratios, and movable-element positions, subject to the transmit-power, quality-of-service, energy-harvesting, energy-splitting, movement-region, and inter-element spacing constraints. The resulting formulation is highly non-convex because the movable-element positions affect the cascaded channels non-linearly and are tightly coupled with the remaining design variables. To obtain a tractable solution, the joint design is decomposed into active beamforming, passive beamforming, movable-element positioning, and PS-ratio optimization blocks, which are updated iteratively through suitable convex reformulations. In particular, the element-position block is handled in a sequential manner, where each movable element is optimized using locally tight quadratic surrogates derived from the corresponding first- and second-order channel derivatives. Numerical results show that the proposed framework delivers higher sum rates than the considered benchmark schemes, exhibits robust behavior under CSI uncertainty and residual HIs, and achieves stable convergence across the considered system configurations.
\end{abstract}

\begin{IEEEkeywords} Simultaneous wireless information and power transfer (SWIPT), movable-element simultaneously transmitting and reflecting surface (ME-STARS), rate-splitting multiple access (RSMA), robust transmission design, hardware impairments.
\end{IEEEkeywords}

\IEEEpeerreviewmaketitle

%\vspace{0.5cm}

\section{Introduction}
\IEEEPARstart{T} {he} rapid growth of connected devices and emerging data-intensive applications is placing increasingly stringent demands on future wireless networks, not only in terms of high data rates, massive connectivity, and reliable communication, but also in providing sustainable energy support for energy-constrained devices \cite{wang2022gcwcn,nguyen20216g}. Meeting these requirements is particularly important for next-generation Internet-of-Things (IoT) networks, where information transmission and energy availability must be addressed simultaneously. In this context, simultaneous wireless information and power transfer (SWIPT) has emerged as a promising approach that enables radio-frequency (RF) signals to carry information while simultaneously delivering wireless energy to the intended devices \cite{zeng2017communications}. However, in multiuser SWIPT networks, inter-user interference can severely degrade the achievable information rates while complicating the joint allocation of resources for information decoding and energy harvesting \cite{park2014joint}. Rate-splitting multiple access (RSMA) provides an effective means of addressing this issue by dividing user messages into common and private parts, allowing interference to be partially decoded and partially treated as noise \cite{mao2022rate}. This flexible interference-management mechanism, together with the ability to adjust the power assigned to common and private streams, makes RSMA particularly well suited to SWIPT, where communication performance and harvested energy must be jointly considered. Therefore, the integration of RSMA with SWIPT offers a promising framework for simultaneously improving spectral efficiency, interference management, and wireless energy delivery in future energy-sustainable networks. The potential of jointly employing SWIPT and RSMA has attracted increasing research interest across a range of wireless communication scenarios, including cognitive radio (CR) networks \cite{acosta2020joint}, ultra-reliable and low-latency communications (URLLC) \cite{karim2025finite}, and cell-free massive multiple-input multiple-output (CF-mMIMO) systems \cite{galappaththige2024sum}.

Although the integration of RSMA with SWIPT provides greater flexibility in interference management and the allocation of transmit power between common and private streams \cite{asif2026robust}, its performance remains highly dependent on the wireless channel conditions, since severe path loss, signal blockage, and unfavorable propagation can reduce both the achievable information rate and the RF power available for energy harvesting. Reconfigurable intelligent surface (RIS) has emerged as an energy-efficient technology for improving wireless propagation by intelligently controlling the reflected signals \cite{liu2021reconfigurable,asif2025noma}. However, conventional RIS operates only in reflection mode, which generally restricts service to users located on one side of the surface. To overcome this coverage restriction, simultaneously transmitting and reflecting surface (STARS) enables an impinging signal to be divided into transmission and reflection components, allowing users on either side of the surface to be served \cite{mu2021simultaneously,asif2026robust123}. Accordingly, several recent studies have investigated STARS-assisted RSMA-SWIPT systems to exploit the interference-management capability of RSMA and the full-space coverage of STARS for improving both information transmission and wireless energy transfer \cite{hashempour2024secure,amiri2025resource,asif2024leveraging}. Nevertheless, in conventional STARS-assisted RSMA-SWIPT systems, the positions of the STARS elements are fixed, which limits the flexibility of the cascaded channels and restricts the system’s ability to exploit spatial variations in the propagation environment. In this regard, movable antenna (MA) technology has recently attracted considerable attention, as it allows antenna elements to adjust their positions within a predefined region to obtain more favorable channel conditions \cite{zhu2023modeling}. Inspired by this idea, movable-element RIS (ME-RIS) has emerged as a new architecture in which the positions of the reflecting elements can be adjusted together with their reflection coefficients\cite{hokmabadi2026joint,zhou2025movable, hu2024intelligent,zhao2026movable}. Recent research has therefore considered ME-RIS architectures in a variety of wireless communication scenarios. In particular, \cite{hokmabadi2026joint} studied an ME-RIS-assisted full-duplex multiple-input single-output (MISO) architecture, where joint beamforming and element-position design was employed to improve the system sum rate. In \cite{zhou2025movable}, an ME-RIS-assisted communication system was considered, where the element positions and reflection coefficients were jointly designed to enhance the achievable rate. The work in \cite{hu2024intelligent} introduced an ME-RIS design for mitigating phase-distribution mismatch in Rician channels and developed a unified non-uniform phase-shift strategy that improves performance while lowering computational complexity. The authors in \cite{zhao2026movable} exploited the position flexibility of ME-RIS elements to enhance the effective channel and improve the achievable transmission rate.
 
 Following the development of ME-RIS, element mobility has also been incorporated into STARS architectures, leading to ME-STARS \cite{zhu2025movable,zhao2025movable,asif2026exploiting,zhao2026exploiting}, in which both the element positions and the  reflection/transmission coefficients are treated as configurable design variables. Specifically, \cite{zhu2025movable} considered an ME-STARS-enabled near-field  wideband system and optimized the beamforming together with the element positions to mitigate the beam-squint effect. For secure transmission, \cite{zhao2025movable} employed ME-STARS and jointly designed the precoding vectors, STARS coefficients, and movable-element positions to enhance the secrecy performance. In \cite{asif2026exploiting}, a robust ME-STARS-enabled RSMA design was proposed, where the transmit precoders, allocation of the common stream, passive STARS response, and movable-element positions were jointly designed to maximize the achievable sum rate. For multiuser transmission, \cite{zhao2026exploiting} developed an ME-STARS design in which the transmit precoders, passive surface response, and movable-element positions were jointly optimized to maximize the weighted sum rate. Consequently, by simultaneously reshaping the cascaded channels toward users in both transmission and reflection regions, ME-STARS provides additional flexibility that is particularly beneficial for RSMA-SWIPT systems, since it can improve the achievable information rates while increasing the RF power available for energy harvesting.

Despite the progress reported in the above studies, the potential of ME-STARS in RSMA-SWIPT networks remains largely unexplored. Specifically, several limitations remain in the existing literature: \textbf{1)} Prior ME-STARS studies have mainly focused on conventional multiuser communication, secure transmission, near-field wideband systems, and RSMA-based information transmission \cite{zhu2025movable,zhao2025movable,asif2026exploiting,zhao2026exploiting}, while their application to simultaneous information and energy transfer has received limited attention. Consequently, it remains unclear how movable-element positioning can be jointly designed with reflection/transmission control, RSMA beamforming, common-rate allocation, and power-splitting (PS) ratios to improve the achievable information rates and harvested energy while satisfying the quality of service (QoS) and energy-harvesting requirements of SWIPT users; \textbf{2)} Moreover, most existing ME-STARS studies either assume perfect channel state information (CSI) or do not jointly consider CSI uncertainty and residual transceiver hardware impairments (HIs). Specifically, CSI uncertainty is critical in ME-STARS-assisted RSMA-SWIPT systems because the cascaded base station (BS)--ME-STARS--user channels vary with the movable-element positions. Therefore, CSI errors can influence the active precoders and ME-STARS passive beamforming, as well as the common-rate allocation, PS ratios, and movable-element positioning, potentially reducing both the achievable information rates and the harvested energy. Moreover, residual HIs introduce a different challenge, as the distortion generated by practical transmitter and receiver components changes the effective desired-signal and interference powers, as well as the total RF power delivered to the energy-harvesting circuit. Consequently, HIs directly affect the decoding SINRs as well as the RF power available for energy harvesting. The combined presence of CSI uncertainty and residual HIs therefore introduces strong coupling among the active beamforming, ME-STARS configuration, common-rate allocation, PS ratios, and movable-element positions, thereby substantially complicating the resulting robust optimization. Thus, a robust ME-STARS-enabled RSMA-SWIPT design that jointly considers CSI uncertainty and residual HIs while optimizing the active precoders, ME-STARS passive beamforming, common-rate allocation, PS ratios, and movable-element positions has not yet been reported in the literature. Accordingly, the key contributions of this study are outlined below:

\begin{itemize}
	\item We develop a robust RSMA-SWIPT transmission design enabled by ME-STARS that jointly exploits movable-element positioning, full-space reflection/transmission control, and RSMA-based interference management for simultaneous wireless information and energy transfer. Accordingly, we formulate a robust sum-rate maximization problem by jointly designing the BS precoders, user-specific common-rate allocation, ME-STARS passive beamforming, PS ratios, and the spatial positions of the ME-STARS elements, subject to transmit-power, QoS, energy-harvesting, energy-splitting, movement-region, and minimum inter-element spacing constraints. A practical nonlinear energy-harvesting model is also incorporated to characterize the operation of the SWIPT receivers.
	
	\item To solve the resulting highly coupled non-convex problem, we develop an iterative optimization framework that successively updates the active beamforming, ME-STARS passive beamforming, spatial positions of the movable elements, and PS ratios. The active and passive beamforming subproblems are handled through tractable convex reformulations, whereas the movable elements are optimized sequentially using locally tight quadratic surrogates derived from the first- and second-order channel derivatives with respect to the element positions. The PS ratios are further optimized through a convex reformulation that directly balances the information-decoding and energy-harvesting requirements of the SWIPT users.
	
	\item The proposed design jointly accounts for CSI uncertainty and residual transceiver HIs, both of which directly affect the operation of the considered RSMA-SWIPT system. In particular, CSI uncertainty propagates through the position-dependent cascaded channels and influences the beamforming, common-rate allocation, ME-STARS configuration, PS ratios, and element-position design. Meanwhile, residual HIs modify the effective decoding SINRs and the RF power delivered to the energy-harvesting circuits. The resulting formulation therefore provides a robust transmission design under practical system imperfections.
	
	\item Finally, extensive numerical results demonstrate the advantages of jointly exploiting ME-STARS, RSMA, and SWIPT compared with the considered benchmark schemes. The results also illustrate the benefits of movable-element optimization and quantify how CSI uncertainty and residual HIs affect the system sum rate, while confirming the stable convergence of the proposed algorithm.
\end{itemize}   
 \begin{figure}[!t]
	\centering
	\includegraphics [width=0.50\textwidth]{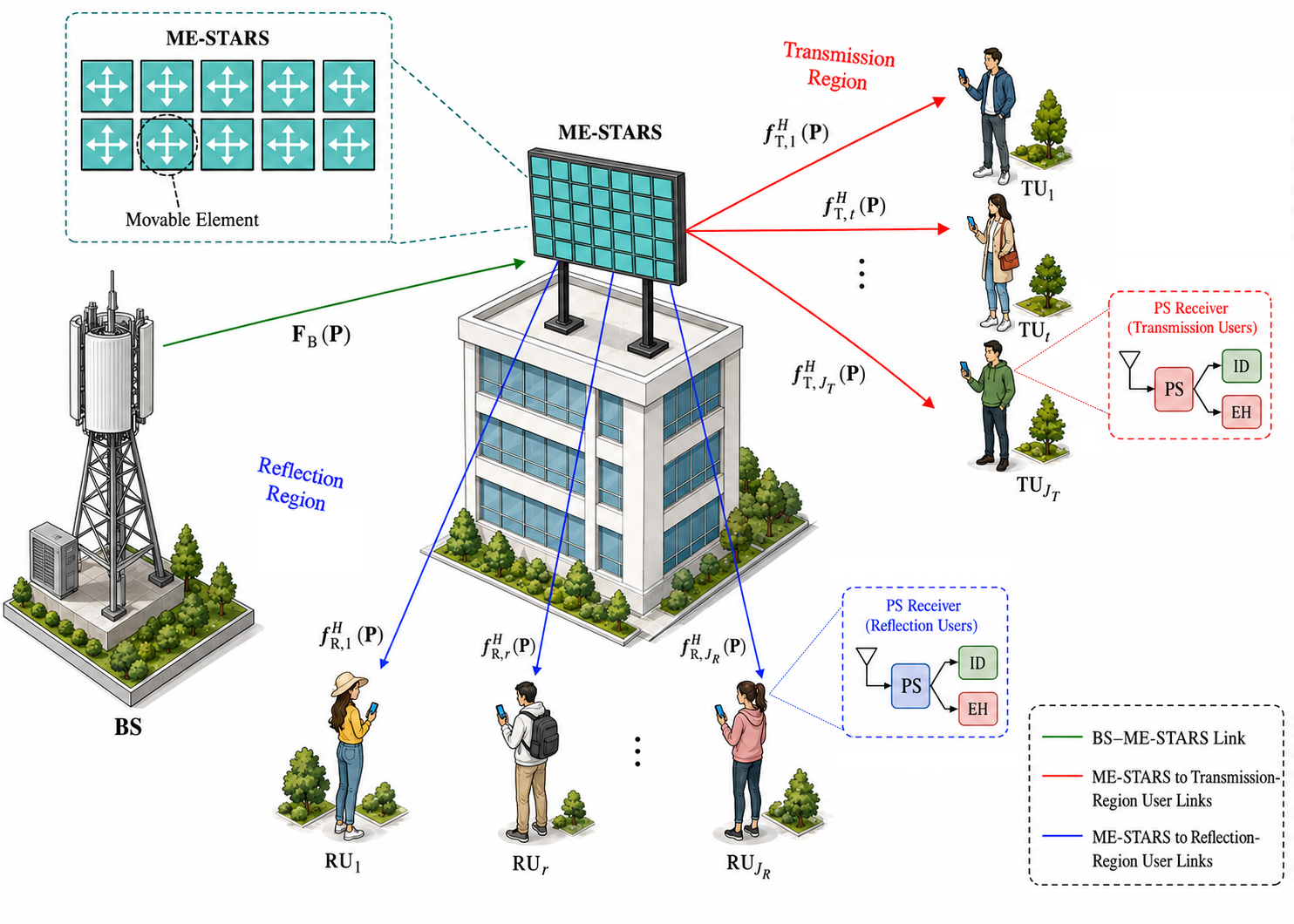}
	\caption{Illustration of the system model.}
	\label{f1}
\end{figure}

\section{System Description and Optimization Framework}
\label{sec:System_Model}
\subsection{RSMA-SWIPT Architecture Enabled by ME-STARS}
\label{subsec:ME_STARS_System}
The downlink RSMA-SWIPT system consists of an $M$-antenna BS, an ME-STARS with $L$ movable elements, and $J$ single-antenna users, as illustrated in Fig.~1. To facilitate simultaneous information and energy transfer, each user adopts a PS receiver that divides the received signal between the information decoding (ID) and energy harvesting (EH) circuits \cite{asif2026robust}. Let $\mathcal J_R=\{1,\ldots,J_R\}$ and $\mathcal J_T=\{1,\ldots,J_T\}$ denote the index sets of users in the reflection and transmission regions, respectively. To differentiate the users in the two regions, we define $\mathcal S_R \triangleq \{(R,r):r\in\mathcal J_R\}$ and $\mathcal S_T \triangleq \{(T,t):t\in\mathcal J_T\}$. The complete user set is then represented by $\mathcal{S}\triangleq \mathcal{S}_{\rm R}\cup\mathcal{S}_{\rm T}$, with $|\mathcal{S}|=J=J_{\rm R}+J_{\rm T}$. Furthermore, owing to severe blockage between the BS and the users, the direct BS--user links are assumed to be unavailable. Hence, both information transmission and wireless energy transfer are realized through the ME-STARS. Each ME-STARS element can move within a prescribed two-dimensional region, providing an additional spatial degree of freedom for system design. Let $\mathbf p_{\ell}=[p_{x,\ell},p_{y,\ell}]^{T}\in\mathbb{R}^{2}$ denote the position of the $\ell$-th movable element, where $\ell\in\mathcal{L}$ and $\mathcal{L}\triangleq\{1,\ldots,L\}$. The overall configuration of the ME-STARS elements is represented by $\mathbf P=[\mathbf p_{1},\mathbf p_{2},\ldots,\mathbf p_{L}]\in\mathbb{R}^{2\times L}$. The position of each movable element, $\mathbf p_\ell$, is confined to a
square region $\mathcal A\subset\mathbb R^2$ with side length $D$, where $\mathbf p_\ell\in\mathcal A$, $\forall \ell\in\mathcal L$.

Moreover, to ensure physical realizability and reduce mutual coupling, any two
distinct movable elements are required to maintain a separation of at least $d_s$, i.e.,
\begin{equation}
\|\mathbf p_{\ell}-\mathbf p_{\ell'}\|_{2}
\geq d_{\rm s},
\qquad
\forall \ell,\ell'\in\mathcal L,\ \ell\neq\ell'.
	\label{eq:minimum_spacing}
\end{equation}

The ME-STARS is assumed to operate according to the energy-splitting (ES) protocol \cite{mu2021simultaneously,asif2026robust123}, under which each incident signal is divided into reflection and
transmission components. The corresponding ME-STARS coefficient matrices are given by
\begin{align}
	\mathbf{\Omega}_{\rm R}
	&=
	\operatorname{diag}
	\left(
	a_{{\rm R},1}e^{j\phi_{{\rm R},1}},
	\ldots,
	a_{{\rm R},L}e^{j\phi_{{\rm R},L}}
	\right),
	\label{eq:Omega_R}\\
	\mathbf{\Omega}_{\rm T}
	&=
	\operatorname{diag}
	\left(
	a_{{\rm T},1}e^{j\phi_{{\rm T},1}},
	\ldots,
	a_{{\rm T},L}e^{j\phi_{{\rm T},L}}
	\right),
	\label{eq:Omega_T}
\end{align}
where $a_{{\rm R},\ell}$ and $a_{{\rm T},\ell}$ denote the reflection and transmission amplitudes of element $\ell$, respectively, while $\phi_{{\rm R},\ell}$ and $\phi_{{\rm T},\ell}$ denote their corresponding phase shifts. The reflection and transmission coefficients under the ES protocol satisfy
\begin{align}
	a_{{\rm R},\ell}^{2}
	+a_{{\rm T},\ell}^{2}
	&\leq1,
	\qquad \forall\ell\in\mathcal{L},
	\label{eq:ES_power}\\
	0\leq a_{{\rm R},\ell},
	a_{{\rm T},\ell}
	&\leq1,
	\qquad \forall\ell\in\mathcal{L},
	\label{eq:ES_amplitude}\\
	0\leq\phi_{{\rm R},\ell},
	\phi_{{\rm T},\ell}
	&<2\pi,
	\qquad \forall\ell\in\mathcal{L}.
	\label{eq:ES_phase}
\end{align}

We next describe the spatial response associated with the movable-element positions. For a propagation direction characterized by azimuth $\theta$ and elevation $\psi$, the ME-STARS steering vector takes the form \cite{zhu2023modeling}:
\begin{equation}
	\mathbf a_{\rm S}(\theta,\psi,\mathbf P)
	=
	\left[
	e^{jk_{\lambda}\delta_{1}
		(\theta,\psi,\mathbf p_{1})},
	\ldots,
	e^{jk_{\lambda}\delta_{L}
		(\theta,\psi,\mathbf p_{L})}
	\right]^{T},
	\label{eq:ME_STARS_steering}
\end{equation}
where $k_\lambda=2\pi/\lambda$ is the wavenumber corresponding to the carrier wavelength $\lambda$. The quantity $\delta_\ell(\theta,\psi,\mathbf p_\ell)$ represents the propagation path difference between the $\ell$-th movable element and the ME-STARS reference point, and is expressed as
\begin{equation}
	\delta_{\ell}(\theta,\psi,\mathbf p_{\ell})
	=
	p_{x,\ell}\sin(\theta)\cos(\psi)
	+
	p_{y,\ell}\sin(\psi).
	\label{eq:path_difference}
\end{equation}
Accordingly, adjusting $\mathbf p_\ell$ changes the phase of the channel component associated with the $\ell$-th element, thereby providing an additional spatial degree of freedom beyond conventional reflection/transmission coefficient design.

Moreover, assuming that the BS is equipped with a uniform linear array with inter-antenna spacing $\lambda/2$, the steering vector associated with departure angle $\theta_B$ is given by
\begin{equation}
	\mathbf a_{\rm B}(\theta_{\rm B})
	=
	\left[
	1,
	e^{j\pi\sin(\theta_{\rm B})},
	\ldots,
	e^{j(M-1)\pi\sin(\theta_{\rm B})}
	\right]^{T}.
	\label{eq:BS_steering}
\end{equation}

Under the far-field Rician channel model, the BS--ME-STARS link is represented as
\begin{equation}
	\begin{aligned}
		\mathbf F_{\rm B}(\mathbf P)
		=\sqrt{\beta_{\rm B}}\Bigg[
		&\sqrt{\frac{\rho_{\rm B}}{\rho_{\rm B}+1}}
		\mathbf a_{\rm S}(\theta_{\rm A},\psi_{\rm A},\mathbf P)
		\mathbf a_{\rm B}^{H}(\theta_{\rm B})\\
		&+\sqrt{\frac{1}{\rho_{\rm B}+1}}
		\mathbf W_{\rm B}(\mathbf P)
		\Bigg],
	\end{aligned}
	\label{eq:BS_ME_STARS_channel}
\end{equation}
where $\beta_{\rm B}$ represents the large-scale gain of the BS--ME-STARS channel, while $\rho_{\rm B}$ denotes its Rician factor. Here, $\theta_{\rm A}$ and $\psi_{\rm A}$ represent the azimuth and elevation of arrival of the line-of-sight (LoS) path at the ME-STARS, while $\theta_{\rm B}$ is its departure angle at the BS. The position-dependent non-line-of-sight (NLoS) component $\mathbf W_{\rm B}(\mathbf P)\in\mathbb C^{L\times M}$
is modeled as
\begin{equation}
	\mathbf W_{\rm B}(\mathbf P)
	=
	\frac{1}{\sqrt{R_{\rm B}}}
	\sum_{r=1}^{R_{\rm B}}
	g_r\,
	\mathbf a_{\rm S}(\theta_r,\psi_r,\mathbf P)
	\mathbf a_{\rm B}^{H}(\vartheta_r),
	\label{eq:BS_STARS_NLoS}
\end{equation}
where $R_{\rm B}$ denotes the number of NLoS paths, $g_r\sim\mathcal{CN}(0,1)$ are independent complex path gains, $(\theta_r,\psi_r)$ are their arrival angles at the ME-STARS, and $\vartheta_r$ is their departure angle at the BS. The factor $1/\sqrt{R_{\rm B}}$ ensures unit average power per NLoS channel entry. The path gains and angles remain fixed within each channel coherence block, while the spatial phases vary with the element positions.

The ME-STARS--user links are assumed to experience comparatively short propagation distances and dominant LoS conditions. Accordingly, the channels associated with the reflection- and transmission-region users are modeled as \cite{liu2026joint}:
\begin{align}
	\mathbf f_{{\rm R},r}^{H}(\mathbf P)
	&=
	\sqrt{\gamma_{{\rm R},r}}
	\mathbf a_{\rm S}^{H}
	(\theta_{{\rm R},r},
	\psi_{{\rm R},r},
	\mathbf P),
	\quad
	r\in\mathcal J_{\rm R},
	\label{eq:R_user_channel}\\
	\mathbf f_{{\rm T},t}^{H}(\mathbf P)
	&=
	\sqrt{\gamma_{{\rm T},t}}
	\mathbf a_{\rm S}^{H}
	(\theta_{{\rm T},t},
	\psi_{{\rm T},t},
	\mathbf P),
	\quad
	t\in\mathcal J_{\rm T},
	\label{eq:T_user_channel}
\end{align}
where $\gamma_{{\rm R},r}$ and $\gamma_{{\rm T},t}$ denote the corresponding large-scale channel gains.

Next, for a generic user $\nu\in\mathcal S$, we define
\begin{equation}
	(\mathbf f_{\nu},\mathbf\Omega_{\nu})
	=
	\begin{cases}
		(\mathbf f_{{\rm R},r},\mathbf\Omega_{\rm R}),
		& \nu=({\rm R},r),\\[1mm]
		(\mathbf f_{{\rm T},t},\mathbf\Omega_{\rm T}),
		& \nu=({\rm T},t).
	\end{cases}
	\label{eq:generic_user_definition}
\end{equation}
Hence, the estimated cascaded BS--ME-STARS--user channel is expressed as
\begin{equation}
	\widehat{\mathbf z}_{\nu}(\mathbf P)
	=
	\mathbf f_{\nu}^{H}(\mathbf P)
	\mathbf\Omega_{\nu}^{H}
	\mathbf F_{\rm B}(\mathbf P)
	\in\mathbb C^{1\times M},
	\qquad \nu\in\mathcal S.
	\label{eq:cascaded_channel}
\end{equation}
The corresponding actual cascaded channel is denoted by $\mathbf z_{\nu}(\mathbf P)$, while the discrepancy between $\mathbf z_{\nu}(\mathbf P)$ and its estimate $\widehat{\mathbf z}_{\nu}(\mathbf P)$ is characterized later through the adopted bounded CSI uncertainty model.

\subsection{RSMA Transmission, Hardware Impairments, and SWIPT}
\label{subsec:signal_model}
Under the RSMA framework, the messages intended for the $J$ users are divided into common and private parts. The common message portions are jointly encoded into a common stream $s_{\rm c}$, whereas the private message of user $\nu$ is encoded into an independent stream $s_{\nu}$. Accordingly, the information-bearing transmit signal generated at the BS is
\begin{equation}
	\mathbf x_{\rm I}
	=
	\mathbf w_{\rm c}s_{\rm c}
	+
	\sum_{\nu\in\mathcal S}
	\mathbf w_{\nu}s_{\nu},
	\label{eq:RSMA_signal}
\end{equation}
where $\mathbf w_c\in\mathbb C^{M\times1}$ and $\mathbf w_\nu\in\mathbb C^{M\times1}$ are the precoders associated with the common stream and the private stream of user $\nu$, respectively. The transmitted streams are mutually independent and have unit average power, i.e., $\mathbb E\{|s_c|^2\}=1$ and $\mathbb E\{|s_\nu|^2\}=1$, $\forall \nu\in\mathcal S$.

To account for residual distortion introduced by non-ideal RF components at the BS, the actual transmitted signal is modeled as
\begin{equation}
	\mathbf x
	=
	\mathbf x_{\rm I}
	+
	\boldsymbol\xi_{\rm tx},
	\label{eq:actual_transmit_signal}
\end{equation}
where $\boldsymbol\xi_{\rm tx}$ denotes the residual transmitter distortion noise. For notational convenience, we define $\mathbf X_{\rm c}\triangleq\mathbf w_{\rm c}\mathbf w_{\rm c}^{H}$ and $\mathbf X_{\nu}\triangleq\mathbf w_{\nu}\mathbf w_{\nu}^{H}$, $\forall\nu\in\mathcal S$, as the covariance matrices of the common and private streams, respectively. We further define $\mathbf X_p\triangleq\sum_{\nu\in\mathcal S}\mathbf X_\nu$ as the combined private-stream covariance and $\mathbf X\triangleq\mathbf X_c+\mathbf X_p$
as the covariance matrix of the information-bearing signal. Moreover, the residual distortion at the transmitter is modeled as a zero-mean circularly symmetric complex Gaussian random vector whose covariance is
proportional to the per-antenna transmit power \cite{asif2026robust123,shen2020beamforming,zhang2023robust}, i.e.,
\begin{equation}
	\boldsymbol\xi_{\rm tx}
	\sim
	\mathcal{CN}
	\left(
	\mathbf 0,
	\kappa_{\rm tx}
	\operatorname{Diag}
	\left(
	\operatorname{diag}(\mathbf X)
	\right)
	\right),
	\label{eq:Tx_HI}
\end{equation}
where $\kappa_{\rm tx}\in(0,1)$ represents the residual transmitter hardware impairment level \cite{zhang2023robust}. For subsequent derivations, the corresponding distortion covariance matrix is denoted by $\mathbf D_{\rm tx}\triangleq\kappa_{\rm tx}\operatorname{Diag}\!\left(\operatorname{diag}(\mathbf X)\right)$.

Accordingly, the signal at the input of the PS circuit of user $\nu$ is expressed as
\begin{equation}
	\widetilde y_{\nu}
	=
	\mathbf z_{\nu}(\mathbf P)\mathbf x
	+
	n_{\nu},
	\qquad \nu\in\mathcal S,
	\label{eq:undistorted_received}
\end{equation}
where $n_{\nu}\sim\mathcal{CN}(0,\sigma_{{\rm a},\nu}^{2})$ denotes the thermal noise at the receive antenna. Further, each user employs a PS receiver to simultaneously perform ID and EH. Let $\varrho_{\nu}\in[0,1]$ denote the fraction of the received power allocated to the ID circuit, while the remaining fraction $1-\varrho_{\nu}$ is directed to the EH circuit.

The residual receiver distortion is introduced in the ID branch after power splitting and is modeled as
\begin{equation}
	\xi_{{\rm rx},\nu}
	\sim
	\mathcal{CN}
	\left(
	0,
	\kappa_{{\rm rx},\nu}\varrho_{\nu}
	\mathbb E
	\left\{
	|\widetilde y_{\nu}|^{2}
	\right\}
	\right),
	\label{eq:Rx_HI}
\end{equation}
where $\kappa_{{\rm rx},\nu}\in(0,1)$ characterizes the residual hardware impairment level of user $\nu$ \cite{shen2020beamforming,zhang2023robust}. Accordingly, the signals at the ID and EH branches are respectively given by
\begin{align}
	y_{\nu}^{\rm ID}
	&=
	\sqrt{\varrho_{\nu}}\,
	\widetilde y_{\nu}
	+
	\xi_{{\rm rx},\nu}
	+
	n_{{\rm d},\nu},
	\label{eq:ID_signal}\\
	y_{\nu}^{\rm EH}
	&=
	\sqrt{1-\varrho_{\nu}}\,
	\widetilde y_{\nu},
	\label{eq:EH_signal}
\end{align}
where $n_{{\rm d},\nu}\sim \mathcal{CN}(0,\sigma_{{\rm d},\nu}^{2})$ denotes the additional noise introduced by the ID circuit.

\subsection{Robust Rate Characterization Under CSI Errors}
\label{subsec:robust_rate}

In practice, the cascaded BS--ME-STARS--user channels cannot be estimated perfectly. For each $\nu\in\mathcal S$, we define the actual and estimated channel outer-product matrices as $\mathbf C_{\nu}(\mathbf P)\triangleq \mathbf z_{\nu}^{H}(\mathbf P)\mathbf z_{\nu}(\mathbf P)$ and $\widehat{\mathbf C}_{\nu}(\mathbf P)\triangleq \widehat{\mathbf z}_{\nu}^{H}(\mathbf P)\widehat{\mathbf z}_{\nu}(\mathbf P)$, respectively. To account for channel estimation errors, we adopt a deterministic spectral-norm-bounded uncertainty model \cite{asif2026robust,zhang2023robust,yao2023robust}, under which
\begin{align}
	&\mathbf{C}_{\nu}(\mathbf{P})=
	\widehat{\mathbf{C}}_{\nu}(\mathbf{P})
	+\mathbf{E}_{\nu}, \quad \mathbf{C}_{\nu}(\mathbf{P}) \succeq \mathbf{0},\nonumber \\
	&\mathbf{E}_{\nu}=
	\mathbf{E}_{\nu}^{H}, \quad
	\|\mathbf{E}_{\nu}\|_{2}
	\leq \varepsilon_{\nu}.
	\label{eq:CSI_uncertainty}
\end{align}
Here, $\mathbf{E}_\nu$ denotes the channel uncertainty matrix, and $\hat{\varepsilon}\in[0,1)$ represents the relative uncertainty level of the channel outer-product matrix. We define the uncertainty radius as $\varepsilon_\nu\triangleq\hat{\varepsilon}\|\widehat{\mathbf{C}}_\nu(\mathbf{P})\|_2=\hat{\varepsilon}\|\widehat{\mathbf{z}}_\nu(\mathbf{P})\|_2^2$ \cite{yao2023robust}, where the equality follows from $\widehat{\mathbf{C}}_\nu(\mathbf{P})=\widehat{\mathbf{z}}_\nu^H(\mathbf{P})\widehat{\mathbf{z}}_\nu(\mathbf{P})$. Although $\hat{\varepsilon}$ is fixed, $\varepsilon_\nu$ depends on the element positions and ME-STARS coefficients through the estimated cascaded channel. These dependencies are suppressed in the notation $\varepsilon_\nu$ for brevity. Additionally, for any positive semidefinite matrix $\mathbf Y\succeq 0$, the spectral--nuclear norm duality gives \cite{asif2026robust,zhang2023robust}:
\begin{equation}
	\left|
	\operatorname{Tr}
	\left(
	\mathbf E_{\nu}\mathbf Y
	\right)
	\right|
	\leq
	\|\mathbf E_{\nu}\|_{2}
	\|\mathbf Y\|_{*}
	\leq
	\varepsilon_{\nu}
	\operatorname{Tr}(\mathbf Y),
	\label{eq:norm_duality}
\end{equation}
where $\|\mathbf Y\|_*$ denotes the nuclear norm associated with $\mathbf Y$. Since $\mathbf Y$ is positive semidefinite, its singular values coincide with its non-negative eigenvalues, and hence $\|\mathbf Y\|_{*}=\operatorname{Tr}(\mathbf Y)$.

Accordingly, we define $\mathbf C_{\nu}^{-}(\mathbf{P})\triangleq\widehat{\mathbf C}_{\nu}(\mathbf{P})-\varepsilon_{\nu}\mathbf I_{M}$ and $\mathbf C_{\nu}^{+}(\mathbf{P})\triangleq\widehat{\mathbf C}_{\nu}(\mathbf{P})+\varepsilon_{\nu}\mathbf I_{M}$. Then, the received-power expression under channel uncertainty satisfies
\begin{equation}
	\operatorname{Tr}
	(\mathbf C_{\nu}^{-}(\mathbf{P})\mathbf Y)
	\leq
	\operatorname{Tr}
	(\mathbf C_{\nu}(\mathbf{P})\mathbf Y)
	\leq
	\operatorname{Tr}
	(\mathbf C_{\nu}^{+}(\mathbf{P})\mathbf Y).
	\label{eq:received_power_bounds}
\end{equation}

To derive the robust decoding rates in the presence of residual HIs, we first define $\mathbf X_{-\nu}\triangleq \mathbf X_{\rm p}-\mathbf X_{\nu}, \nu\in\mathcal S$. During common-stream decoding, the private streams are regarded as interference at user $\nu$. Accounting for the transmitter and receiver distortions, define the corresponding aggregate interference-distortion covariance matrix as
\begin{equation}
	\begin{aligned}
		\mathbf J_{{\rm c},\nu}
		\triangleq\,
		&(1+\kappa_{{\rm rx},\nu})\mathbf X_{\rm p}
		+
		\kappa_{{\rm rx},\nu}\mathbf X_{\rm c} +
		(1+\kappa_{{\rm rx},\nu})\mathbf D_{\rm tx}.
	\end{aligned}
	\label{eq:J_common}
\end{equation}
A conservative lower bound on the worst-case SINR for decoding the common stream is given by
\begin{equation}
	\Gamma_{{\rm c},\nu}^{\rm wc}
	=
	\frac{
		\varrho_{\nu}
		\operatorname{Tr}
		\left(
		\mathbf C_{\nu}^{-}(\mathbf{P})\mathbf X_{\rm c}
		\right)}
	{
		\varrho_{\nu}
		\operatorname{Tr}
		\left(
		\mathbf C_{\nu}^{+}(\mathbf{P})\mathbf J_{{\rm c},\nu}
		\right)
		+
		\sigma_{{\rm ID},\nu}^{2}
	},
	\qquad \nu\in\mathcal S,
	\label{eq:robust_common_SINR}
\end{equation}
where $\sigma_{{\rm ID},\nu}^{2}=\varrho_{\nu}(1+\kappa_{{\rm rx},\nu})\sigma_{{\rm a},\nu}^{2}+\sigma_{{\rm d},\nu}^{2}$.

After successfully decoding and removing the common information stream, user $\nu$ proceeds to decode its own private
stream while treating the private streams of the other users as interference. Further, we define
\begin{equation}
	\begin{aligned}
		\mathbf J_{{\rm p},\nu}
		\triangleq\,
		&\scalemath{0.98}{\mathbf X_{-\nu}
		+
		\kappa_{{\rm rx},\nu}\mathbf X_{\rm p}
		+
		\kappa_{{\rm rx},\nu}\mathbf X_{\rm c}+
		(1+\kappa_{{\rm rx},\nu})\mathbf D_{\rm tx}.}
	\end{aligned}
	\label{eq:J_private}
\end{equation}
Accordingly, a conservative lower bound on the corresponding worst-case private-stream SINR is given by
\begin{equation}
	\Gamma_{{\rm p},\nu}^{\rm wc}
	=
	\frac{
		\varrho_{\nu}
		\operatorname{Tr}
		\left(
		\mathbf C_{\nu}^{-}(\mathbf{P})\mathbf X_{\nu}
		\right)}
	{
		\varrho_{\nu}
		\operatorname{Tr}
		\left(
		\mathbf C_{\nu}^{+}(\mathbf{P})\mathbf J_{{\rm p},\nu}
		\right)
		+
		\sigma_{{\rm ID},\nu}^{2}
	},
	\qquad \nu\in\mathcal S.
	\label{eq:robust_private_SINR}
\end{equation}

Accordingly, the robust achievable common- and private-stream rates at user $\nu$ are given by
\begin{align}
	R_{{\rm c},\nu}^{\rm wc}
	&=
	\log_{2}
	\left(
	1+\Gamma_{{\rm c},\nu}^{\rm wc}
	\right),
	\label{eq:robust_common_rate}\\
	R_{{\rm p},\nu}^{\rm wc}
	&=
	\log_{2}
	\left(
	1+\Gamma_{{\rm p},\nu}^{\rm wc}
	\right).
	\label{eq:robust_private_rate}
\end{align}

For each user $\nu$, let $\zeta_{\nu}\geq0$ represent its assigned share of the common rate. The total common rate is therefore
\begin{equation}
	R_{\rm c}^{\rm tot}
	\triangleq
	\sum_{\nu\in\mathcal S}\zeta_{\nu}.
	\label{eq:total_common_rate}
\end{equation}
Because every user is required to decode the common stream, $R_{\rm c}^{\rm tot}$ is limited by the common-stream rate achievable at each user, i.e.,
\begin{equation}
	R_{\rm c}^{\rm tot}
	\leq
	R_{{\rm c},\nu}^{\rm wc},
	\qquad
	\forall\nu\in\mathcal S.
	\label{eq:common_decoding_condition}
\end{equation}
Consequently, the robust achievable rate of user $\nu$ is
\begin{equation}
	R_{\nu}^{\rm wc}
	=
	\zeta_{\nu}
	+
	R_{{\rm p},\nu}^{\rm wc}.
	\label{eq:user_total_rate}
\end{equation}

\subsection{Nonlinear Energy Harvesting Model}
\label{subsec:nonlinear_EH}
We next characterize the RF power delivered to the EH circuit. Unlike the ID circuit, the EH branch collects energy from the desired signal, inter-user interference, and transmitter distortion. For notational convenience, define
\begin{equation}
	\mathbf J_{{\rm EH},\nu}
	\triangleq
	\left(
	\mathbf X+\mathbf D_{\rm tx}
	\right).
	\label{eq:J_EH}
\end{equation}
Accordingly, under the considered bounded CSI uncertainty, a guaranteed lower bound on the RF power supplied to the EH circuit is given by
\begin{equation}
	\begin{aligned}
		P_{{\rm RF},\nu}^{\rm lb}
		=
		(1-\varrho_{\nu})
		\Big[
		&\operatorname{Tr}
		\left(
		\mathbf C_{\nu}^{-}
		\mathbf J_{{\rm EH},\nu}
		\right)+
		\sigma_{{\rm a},\nu}^{2}
		\Big].
	\end{aligned}
	\label{eq:robust_RF_power}
\end{equation}

To capture the saturation behavior of practical rectifying circuits, we adopt a nonlinear logistic EH model \cite{boshkovska2015practical,xiong2017rate}. Specifically, the harvested DC power at user $\nu$ is expressed as
\begin{equation}
	P_{{\rm EH},\nu}
	=
	\frac{
		\mathcal{F}_{\nu}
		\left(
		P_{{\rm RF},\nu}^{\rm lb}
		\right)
		-
		P_{\nu}^{\rm sat}\vartheta_{\nu}
	}
	{1-\vartheta_{\nu}},
	\qquad \nu\in\mathcal S,
	\label{eq:nonlinear_EH}
\end{equation}
where
\begin{equation}
	\mathcal{F}_{\nu}\!\left(P_{{\rm RF},\nu}^{\rm lb}\right)
	=
	\frac{
		P_{\nu}^{\rm sat}
	}{
		1+\exp\!\left[
		-a_{\nu}
		\left(
		P_{{\rm RF},\nu}^{\rm lb}-b_{\nu}
		\right)
		\right]
	},
	\label{eq:logistic_function}
\end{equation}
and
\begin{equation}
	\vartheta_{\nu}
	=
	\frac{1}
	{1+\exp(a_{\nu}b_{\nu})}.
	\label{eq:EH_offset}
\end{equation}
Here, $P_{\nu}^{\rm sat}$ denotes the maximum harvested DC
power of the EH circuit, whereas $a_{\nu}$ and $b_{\nu}$ are
circuit-dependent parameters characterizing its nonlinear
RF-to-DC conversion behavior.

\subsection{Problem Formulation}
\label{subsec:problem_formulation}
We aim to maximize the robust achievable sum rate through the joint design of the BS precoders, user-specific common-rate allocation, ME-STARS passive beamforming, PS ratios, and the spatial positions of the movable elements.
\begin{subequations}
	\label{prob:P1}
	\begin{align}
		\mathrm{(P1)}\quad
		\max_{\mathcal V}\quad
		&
		\sum_{\nu\in\mathcal S}
		\left(
		\zeta_{\nu}
		+
		R_{{\rm p},\nu}^{\rm wc}
		\right)
		\label{prob:P1a}\\
		\mathrm{s.t.}\quad
		&
		\sum_{\nu\in\mathcal S}\zeta_{\nu}
		\leq
		R_{{\rm c},\nu}^{\rm wc},
		\quad
		\forall\nu\in\mathcal S,
		\label{prob:P1b}\\
		&
		\zeta_{\nu}
		+
		R_{{\rm p},\nu}^{\rm wc}
		\geq
		R_{\nu}^{\min},
		\quad
		\forall\nu\in\mathcal S,
		\label{prob:P1c}\\
		&
		P_{{\rm EH},\nu}
		\geq
		P_{{\rm EH},\nu}^{\min},
		\quad
		\forall\nu\in\mathcal S,
		\label{prob:P1d}\\
		&
		\scalemath{0.90}{(1+\kappa_{\rm tx})
		\Big(\|\mathbf w_{\rm c}\|_2^2
		+\sum_{\nu\in\mathcal S}\|\mathbf w_\nu\|_2^2\Big)
		\leq P_{\rm tx}^{\max},}
		\label{prob:P1e}\\
		&
		0\leq\varrho_{\nu}\leq1,
		\quad
		\forall\nu\in\mathcal S,
		\label{prob:P1f}\\
		&
		\mathbf p_{\ell}\in\mathcal A,
		\quad
		\forall\ell\in\mathcal L,
		\label{prob:P1g}\\
		&
		\|\mathbf p_{\ell}-\mathbf p_{\ell'}\|_{2}
		\geq
		d_{\rm s},
		\quad
		\forall \ \ell\neq\ell',
		\label{prob:P1h}\\
		&
		a_{{\rm R},\ell}^{2}
		+
		a_{{\rm T},\ell}^{2}
		\leq1,
		\quad
		\forall\ell\in\mathcal L,
		\label{prob:P1i}\\
		&
		0\leq
		a_{{\rm R},\ell},
		a_{{\rm T},\ell}
		\leq1,
		\quad
		\forall\ell\in\mathcal L,
		\label{prob:P1j}\\
		&
		0\leq
		\phi_{{\rm R},\ell},
		\phi_{{\rm T},\ell}
		<2\pi,
		\quad
		\forall\ell\in\mathcal L,
		\label{prob:P1k}\\
		&
		\zeta_{\nu}\geq0,
		\quad
		\forall\nu\in\mathcal S.
		\label{prob:P1l}
	\end{align}
\end{subequations}
where $\mathcal{V}\triangleq\Big\{\mathbf w_{\rm c},\{\mathbf w_{\nu}\}_{\nu\in\mathcal S}, \{\zeta_{\nu}\}_{\nu\in\mathcal S},\{\varrho_{\nu}\}_{\nu\in\mathcal S},\mathbf\Omega_{\rm R},\mathbf\Omega_{\rm T},\mathbf P\Big\}$. Constraint \eqref{prob:P1b} guarantees that the common stream can be decoded by every user, whereas \eqref{prob:P1c} imposes the individual communication QoS requirements. Constraint \eqref{prob:P1d} ensures that the harvested DC power at each user satisfies the prescribed minimum requirement. Constraint \eqref{prob:P1e} limits the total radiated power,
including both the information-bearing signal and transmitter distortion, where $P_{\rm tx}^{\max}$
denotes the BS total radiated-power budget, while \eqref{prob:P1f} specifies the feasible range of the power-splitting ratios. Moreover, \eqref{prob:P1g} and \eqref{prob:P1h} restrict the movable elements to the allowable region while maintaining the required inter-element separation. Finally, \eqref{prob:P1i}--\eqref{prob:P1k} specify the feasible reflection/transmission operation of the ME-STARS under the ES protocol.

Problem \eqref{prob:P1} remains highly non-convex because the BS precoders, allocation of the common rate, PS ratios, ME-STARS passive beamforming, and spatial positions of the movable elements are tightly coupled. In particular, the cascaded channels depend non-linearly on the spatial coordinates of the movable elements, whereas the transmitter and receiver distortion powers are coupled with the optimization variables through the corresponding transmit and received signal powers. The robust common- and private-stream rates further involve non-convex fractional expressions, and the non-linear EH model couples the information- and energy-transfer designs through the power-splitting variables. In addition, the minimum inter-element spacing constraints are non-convex. These challenges motivate an iterative optimization framework, developed in the following section, where the BS precoders, ME-STARS passive beamforming, PS ratios, and spatial positions of the movable elements are successively updated.

\section{Iterative Optimization Design}

\subsection{Transmit Beamforming Optimization}
\label{subsec:active_beamforming}
We first focus on the active beamforming design by fixing the ME-STARS coefficients, movable-element positions, and PS ratios. Accordingly, the position-dependent channel matrices $\mathbf C_{\nu}^{-}(\mathbf P)$ and $\mathbf C_{\nu}^{+}(\mathbf P)$ remain fixed during this update. For notational simplicity, the argument $\mathbf P$ is omitted throughout this subsection. Problem \eqref{prob:P1} then reduces to
\begin{subequations}
	\label{prob:P2}
	\begin{align}
		\mathrm{(P2)}\quad
		\max_{\mathcal V_{\rm A}}\quad
		&
		\sum_{\nu\in\mathcal S}
		\left(
		\zeta_{\nu}
		+
		R_{{\rm p},\nu}^{\rm wc}
		\right)
		\label{prob:P2a}\\
		\mathrm{s.t.}\quad
		& \eqref{prob:P1b} -\eqref{prob:P1e}, \eqref{prob:P1l}. \label{prob:P2b}
	\end{align}
\end{subequations}
where $\mathcal V_{\rm A}\triangleq\left\{\mathbf w_{\rm c},\{\mathbf w_{\nu}\}_{\nu\in\mathcal S},\{\zeta_{\nu}\}_{\nu\in\mathcal S}\right\}$. Since the ME-STARS coefficients, element positions, and PS ratios are fixed in this subproblem, the estimated channel matrices $\widehat{\mathbf{C}}_\nu(\mathbf{P})$ and the corresponding uncertainty radii $\varepsilon_\nu =\hat{\varepsilon}\|\widehat{\mathbf{C}}_\nu(\mathbf{P})\|_2$ are fixed during the transmit-beamforming update. Nevertheless, problem \eqref{prob:P2} remains non-convex because of the coupled robust rate expressions, the signal-dependent distortion terms, and the nonlinear EH constraint.

To facilitate a tractable reformulation of problem \eqref{prob:P2}, the precoding vectors are lifted to the positive semidefinite covariance matrices $\mathbf X_{\rm c}=\mathbf w_{\rm c}\mathbf w_{\rm c}^{H}$ and
$\mathbf X_{\nu}=\mathbf w_{\nu}\mathbf w_{\nu}^{H}$, $\forall\nu\in\mathcal S$. These matrices satisfy
\begin{align}
	&\mathbf X_{\rm c}\succeq\mathbf0,
	\qquad
	\operatorname{rank}(\mathbf X_{\rm c})\leq1,
	\label{eq:AB_Xc_rank}\\
	&\mathbf X_{\nu}\succeq\mathbf0,
	\qquad
	\operatorname{rank}(\mathbf X_{\nu})\leq1,
	\quad
	\forall\nu\in\mathcal S.
	\label{eq:AB_Xnu_rank}
\end{align}

 Further, the transmitter-distortion covariance can be expressed in terms of the lifted covariance matrices as
\begin{equation}
	\mathbf D_{\rm tx}
	=
	\kappa_{\rm tx}
	\operatorname{Diag}
	\left[
	\operatorname{diag}
	\left(
	\mathbf X_{\rm c}
	+
	\sum_{\nu\in\mathcal S}
	\mathbf X_{\nu}
	\right)
	\right],
	\label{eq:AB_Dtx}
\end{equation}
which is linear with respect to the transmit covariance matrices.

We first simplify the nonlinear harvested-power constraint in \eqref{prob:P1d}. Since the logistic EH function in \eqref{eq:logistic_function} is strictly increasing with respect to its RF input power, \eqref{prob:P1d} can be equivalently rewritten as
\begin{equation}
	\mathcal F_{\nu}
	\left(
	P_{{\rm RF},\nu}^{\rm lb}
	\right)
	\geq
	\chi_{\nu},
	\qquad
	\forall\nu\in\mathcal S,
	\label{eq:AB_EH_intermediate}
\end{equation}
where
\begin{equation}
	\chi_{\nu}
	\triangleq
	P_{\nu}^{\rm sat}\vartheta_{\nu}
	+
	(1-\vartheta_{\nu})
	P_{{\rm EH},\nu}^{\min}.
	\label{eq:AB_chi}
\end{equation}
For $0\leq P_{{\rm EH},\nu}^{\min}<P_{\nu}^{\rm sat}$, the inverse of the logistic function yields the equivalent RF-input requirement
\begin{equation}
	P_{{\rm RF},\nu}^{\rm lb}
	\geq
	P_{{\rm RF},\nu}^{\rm req},
	\qquad
	\forall\nu\in\mathcal S,
	\label{eq:AB_EH_RF_constraint}
\end{equation}
where
\begin{equation}
	P_{{\rm RF},\nu}^{\rm req}
	\triangleq
	b_{\nu}
	+
	\frac{1}{a_{\nu}}
	\ln
	\left(
	\frac{\chi_{\nu}}
	{P_{\nu}^{\rm sat}-\chi_{\nu}}
	\right),
	\label{eq:AB_RF_threshold}
\end{equation}
and
\begin{equation}
	\begin{aligned}
		P_{{\rm RF},\nu}^{\rm lb}
		=
		(1-\varrho_{\nu})
		\Big[
		&\operatorname{Tr}
		\left(
		\mathbf C_{\nu}^{-}
		(\mathbf X+\mathbf D_{\rm tx})
		\right)+
		\sigma_{{\rm a},\nu}^{2}
		\Big],
	\end{aligned}
	\label{eq:AB_RF_affine}
\end{equation}
which is affine with respect to the transmit covariance matrices.

To facilitate the subsequent reformulation, we define the quantities associated with common- and private-stream decoding at user $\nu$ as $\mathcal I_{{\rm c},\nu}\triangleq \varrho_{\nu}\operatorname{Tr}\!\left(\mathbf C_{\nu}^{+}\mathbf J_{{\rm c},\nu}\right)+\sigma_{{\rm ID},\nu}^{2}$, $\mathcal T_{{\rm c},\nu}\triangleq \mathcal I_{{\rm c},\nu}+\varrho_{\nu}\operatorname{Tr}\!\left(\mathbf C_{\nu}^{-}\mathbf X_{\rm c}\right)$, $\mathcal I_{{\rm p},\nu}\triangleq \varrho_{\nu}\operatorname{Tr}\!\left(\mathbf C_{\nu}^{+}\mathbf J_{{\rm p},\nu}\right)+\sigma_{{\rm ID},\nu}^{2}$, and $\mathcal T_{{\rm p},\nu}\triangleq \mathcal I_{{\rm p},\nu}+\varrho_{\nu}\operatorname{Tr}\!\left(\mathbf C_{\nu}^{-}\mathbf X_{\nu}\right)$. Here, $\mathcal I_{{\rm c},\nu}$ and $\mathcal I_{{\rm p},\nu}$ denote the corresponding interference-plus-noise terms, whereas $\mathcal T_{{\rm c},\nu}$ and $\mathcal T_{{\rm p},\nu}$ additionally include the desired-signal contribution. Accordingly, the robust rates of the common and private streams can be reformulated as differences of logarithmic terms, i.e.,
\begin{align}
	R_{{\rm c},\nu}^{\rm wc}
	&=
	\log_{2}
	\left(
	\mathcal T_{{\rm c},\nu}
	\right)
	-
	\log_{2}
	\left(
	\mathcal I_{{\rm c},\nu}
	\right).
	\label{eq:AB_Rc_DC}
\end{align}
\begin{align}
	R_{{\rm p},\nu}^{\rm wc}
	&=
	\log_{2}
	\left(
	\mathcal T_{{\rm p},\nu}
	\right)
	-
	\log_{2}
	\left(
	\mathcal I_{{\rm p},\nu}
	\right).
	\label{eq:AB_Rp_DC}
\end{align}

To handle the remaining non-convexity in the rate expressions in \eqref{eq:AB_Rc_DC} and \eqref{eq:AB_Rp_DC}, the second logarithmic term is replaced by a tight first-order upper bound evaluated at the current iterate. Specifically, let $\mathcal I_{{\rm c},\nu}^{(i)}$ and $\mathcal I_{{\rm p},\nu}^{(i)}$ denote the values of
$\mathcal I_{{\rm c},\nu}$ and $\mathcal I_{{\rm p},\nu}$, respectively, at the $i$-th iteration. Owing to the concavity of $\log_2(x)$, a linear approximation obtained from its first-order expansion around the current point provides the following global upper bound:
\begin{equation}
	\log_{2}(x)
	\leq
	\log_{2}
	\left(
	x^{(i)}
	\right)
	+
	\frac{
		x-x^{(i)}
	}{
		x^{(i)}\ln2
	},
	\qquad
	x>0.
	\label{eq:AB_log_upper}
\end{equation}
Hence, a concave lower approximation of the common-stream rate is
given by
\begin{equation}
	\begin{aligned}
		\widetilde R_{{\rm c},\nu}^{(i)}
		\triangleq\,
		&
		\log_{2}
		\left(
		\mathcal T_{{\rm c},\nu}
		\right)
		-
		\log_{2}
		\left(
		\mathcal I_{{\rm c},\nu}^{(i)}
		\right)
		-
		\frac{
			\mathcal I_{{\rm c},\nu}
			-
			\mathcal I_{{\rm c},\nu}^{(i)}
		}{
			\mathcal I_{{\rm c},\nu}^{(i)}
			\ln2
		},
	\end{aligned}
	\label{eq:AB_Rc_lower}
\end{equation}
while the corresponding private-stream approximation is
\begin{equation}
	\begin{aligned}
		\widetilde R_{{\rm p},\nu}^{(i)}
		\triangleq\,
		&
		\log_{2}
		\left(
		\mathcal T_{{\rm p},\nu}
		\right)
		-
		\log_{2}
		\left(
		\mathcal I_{{\rm p},\nu}^{(i)}
		\right)
		-
		\frac{
			\mathcal I_{{\rm p},\nu}
			-
			\mathcal I_{{\rm p},\nu}^{(i)}
		}{
			\mathcal I_{{\rm p},\nu}^{(i)}
			\ln2
		}.
	\end{aligned}
	\label{eq:AB_Rp_lower}
\end{equation}
These approximations satisfy
\begin{equation}
	R_{{\rm c},\nu}^{\rm wc}
	\geq
	\widetilde R_{{\rm c},\nu}^{(i)},
	\qquad
	R_{{\rm p},\nu}^{\rm wc}
	\geq
	\widetilde R_{{\rm p},\nu}^{(i)},
	\quad
	\forall\nu\in\mathcal S,
	\label{eq:AB_rate_lower_bounds}
\end{equation}
with equality at the current successive convex approximation (SCA) point.

We further introduce the auxiliary variables $\tau_{{\rm c},\nu}$ and $\tau_{{\rm p},\nu}$ to represent the
achievable common- and private-stream rates in the convexified problem, respectively. Accordingly,
\begin{align}
	\tau_{{\rm c},\nu}
	&\leq
	\widetilde R_{{\rm c},\nu}^{(i)},
	\quad
	\forall\nu\in\mathcal S.
	\label{eq:AB_tau_c}
\end{align}
\begin{align}
	\tau_{{\rm p},\nu}
	&\leq
	\widetilde R_{{\rm p},\nu}^{(i)},
	\quad
	\forall\nu\in\mathcal S.
	\label{eq:AB_tau_p}
\end{align}

Applying semidefinite relaxation (SDR) to the rank-one constraints in \eqref{eq:AB_Xc_rank} and \eqref{eq:AB_Xnu_rank} yields a tractable relaxation of the active beamforming subproblem, which can be formulated as
\begin{subequations}
	\label{prob:P3}
	\begin{align}
		\mathrm{(P3)}\quad
		\max_{\mathcal Z_{\rm A}}\quad
		&
		\sum_{\nu\in\mathcal S}
		\left(
		\zeta_{\nu}
		+
		\tau_{{\rm p},\nu}
		\right)
		\label{prob:P3a}\\
		\mathrm{s.t.}\quad
		&
		R_{\rm c}^{\rm tot}
		\leq
		\tau_{{\rm c},\nu},
		\quad
		\forall\nu\in\mathcal S,
		\label{prob:P3b}\\
		&
		\zeta_{\nu}
		+
		\tau_{{\rm p},\nu}
		\geq
		R_{\nu}^{\min},
		\quad
		\forall\nu\in\mathcal S,
		\label{prob:P3c}\\
		&
		\tau_{{\rm c},\nu}
		\leq
		\widetilde R_{{\rm c},\nu}^{(i)},
		\quad
		\forall\nu\in\mathcal S,
		\label{prob:P3d}
		\end{align}
		\begin{align}
		&
		\tau_{{\rm p},\nu}
		\leq
		\widetilde R_{{\rm p},\nu}^{(i)},
		\quad
		\forall\nu\in\mathcal S,
		\label{prob:P3e}\\
		&
		P_{{\rm RF},\nu}^{\rm lb}
		\geq
		P_{{\rm RF},\nu}^{\rm req},
		\quad
		\forall\nu\in\mathcal S,
		\label{prob:P3f}\\
		&
		\scalemath{0.87}{(1+\kappa_{\rm tx})
		\Big(
		\operatorname{Tr}(\mathbf X_{\rm c})
		+\sum_{\nu\in\mathcal S}\operatorname{Tr}(\mathbf X_\nu)
		\Big)
		\leq P_{\rm tx}^{\max}},
		\label{prob:P3g}\\
		&
		\mathbf X_{\rm c}\succeq\mathbf0,
		\qquad
		\mathbf X_{\nu}\succeq\mathbf0,
		\quad
		\forall\nu\in\mathcal S,
		\label{prob:P3h}\\
		&
		\mathcal T_{{\rm c},\nu}\geq\epsilon,
		\qquad
		\mathcal T_{{\rm p},\nu}\geq\epsilon,
		\quad
		\forall\nu\in\mathcal S,
		\label{prob:P3i}\\
		&
		\tau_{{\rm c},\nu}\geq0,
		\quad
		\tau_{{\rm p},\nu}\geq0,
		\quad
		\zeta_{\nu}\geq0,
		\quad
		\forall\nu\in\mathcal S,
		\label{prob:P3j}
	\end{align}
\end{subequations}
where $\epsilon>0$ is a sufficiently small constant introduced to ensure that the logarithmic arguments remain strictly positive, and
$\mathcal Z_{\rm A}\triangleq
\left\{
\mathbf X_{\rm c},
\{\mathbf X_{\nu}\}_{\nu\in\mathcal S},
\{\zeta_{\nu}\}_{\nu\in\mathcal S},
\{\tau_{{\rm c},\nu}\}_{\nu\in\mathcal S},
\{\tau_{{\rm p},\nu}\}_{\nu\in\mathcal S}
\right\}$
denotes the set of optimization variables. Problem \eqref{prob:P3} is convex and can therefore be solved efficiently using standard convex solvers. Upon convergence, if the optimized covariance matrices are rank one, the associated precoding vectors can be directly recovered via eigenvalue decomposition; otherwise, feasible active beamforming vectors are obtained through Gaussian randomization \cite{ni2021resource}.

\subsection{ME-STARS Passive Beamforming Design}
\label{subsec:passive_beamforming}
We next focus on the ME-STARS passive beamforming design by fixing the BS precoders, PS ratios, and movable-element positions. The resulting passive beamforming subproblem is formulated as
\begin{subequations}
	\label{prob:P4}
	\begin{align}
		\mathrm{(P4)}\quad
		\max_{\mathcal V_{\rm P}}\quad
		&
		\sum_{\nu\in\mathcal S}
		\left(
		\zeta_{\nu}
		+
		R_{{\rm p},\nu}^{\rm wc}
		\right)
		\label{prob:P4a}
		\\
		\mathrm{s.t.}\quad
		& \eqref{prob:P1b} -\eqref{prob:P1d}, \eqref{prob:P1i}-\eqref{prob:P1l}
		,
		\label{prob:P4h}
	\end{align}
\end{subequations}
where
$\mathcal V_{\rm P}\triangleq
\left\{
\mathbf\Omega_{\rm R},
\mathbf\Omega_{\rm T},
\{\zeta_{\nu}\}_{\nu\in\mathcal S}
\right\}$.
Problem \eqref{prob:P4} remains non-convex because the robust rate and harvested-power expressions depend non-linearly on the reflection/transmission coefficients.

To facilitate a tractable reformulation of the passive beamforming subproblem, we define the reflection and transmission coefficient vectors as $\boldsymbol{\theta}_{\rm R}\triangleq [a_{{\rm R},1}e^{j\phi_{{\rm R},1}},\ldots,a_{{\rm R},L}e^{j\phi_{{\rm R},L}}]^{T}$ and $\boldsymbol{\theta}_{\rm T}\triangleq [a_{{\rm T},1}e^{j\phi_{{\rm T},1}},\ldots,a_{{\rm T},L}e^{j\phi_{{\rm T},L}}]^{T}$, respectively, such that $\mathbf\Omega_{\rm R}=\operatorname{diag}(\boldsymbol{\theta}_{\rm R})$ and $\mathbf\Omega_{\rm T}=\operatorname{diag}(\boldsymbol{\theta}_{\rm T})$. For a generic user $\nu\in\mathcal S$, we further define $\boldsymbol{\theta}_{\nu}\triangleq\boldsymbol{\theta}_{\rm R}$ for $\nu\in\mathcal S_{\rm R}$ and $\boldsymbol{\theta}_{\nu}\triangleq\boldsymbol{\theta}_{\rm T}$ for $\nu\in\mathcal S_{\rm T}$. Further, for notational simplicity, the dependence of the channel matrices on $\mathbf P$ is omitted throughout this subsection. Define $\mathbf G_{\nu}\triangleq \operatorname{diag}(\mathbf f_{\nu})^{H}\mathbf F_{\rm B}$,
$\forall\nu\in\mathcal S$, such that the estimated cascaded channel represented in terms of the passive beamforming coefficients is denoted by $\bar{\mathbf z}_{\nu}\triangleq \boldsymbol{\theta}_{\nu}^{H}\mathbf G_{\nu}$. We further introduce $\mathbf V_{\rm R}\triangleq \boldsymbol{\theta}_{\rm R}\boldsymbol{\theta}_{\rm R}^{H}$ and $\mathbf V_{\rm T}\triangleq \boldsymbol{\theta}_{\rm T}\boldsymbol{\theta}_{\rm T}^{H}$ for the reflection and transmission coefficients, respectively. Further, for compactness, we define
\begin{equation}
	\mathbf V_{\nu}
	\triangleq
	\begin{cases}
		\mathbf V_{\rm R},
		& \nu\in\mathcal S_{\rm R},\\[1mm]
		\mathbf V_{\rm T},
		& \nu\in\mathcal S_{\rm T}.
	\end{cases}
	\label{eq:PB_generic_V}
\end{equation}
Accordingly, the corresponding estimated channel outer-product matrix is denoted by $\bar{\mathbf C}_{\nu}\triangleq \bar{\mathbf z}_{\nu}^{H}\bar{\mathbf z}_{\nu} =\mathbf G_{\nu}^{H}\mathbf V_{\nu}\mathbf G_{\nu}$. By construction, $\mathbf V_{\rm R}\succeq\mathbf0$ and $\mathbf V_{\rm T}\succeq\mathbf0$, with $\operatorname{rank}(\mathbf V_{\rm R})\leq1$ and $\operatorname{rank}(\mathbf V_{\rm T})\leq1$.

Further, since the ME-STARS coefficients are optimized in this subproblem, the uncertainty radius varies with the passive beamforming design. Using $\bar{\mathbf{z}}_\nu=\boldsymbol{\theta}_\nu^H\mathbf{G}_\nu$ and $\mathbf{V}_\nu=\boldsymbol{\theta}_\nu\boldsymbol{\theta}_\nu^H$, we express the uncertainty radius as
\begin{equation}
	\varepsilon_\nu(\mathbf{V}_\nu)
	=
	\hat{\varepsilon}\|\bar{\mathbf{z}}_\nu\|_2^2
	=
	\hat{\varepsilon}
	\operatorname{Tr}\!\left(
	\mathbf{G}_\nu\mathbf{G}_\nu^H\mathbf{V}_\nu
	\right).
	\label{eq:passive_uncertainty_radius}
\end{equation}
This expression explicitly accounts for the dependence of the uncertainty radius on the ME-STARS coefficients.

The lower- and upper-bounding channel matrices associated with the passive beamforming representation are defined as
\begin{equation}
	\bar{\mathbf{C}}_\nu^{-}
	\triangleq
	\bar{\mathbf{C}}_\nu
	- \varepsilon_\nu(\mathbf{V}_\nu)\mathbf{I}_M,
		\label{eq:PB_C_lower}
\end{equation}
and
\begin{equation}
	\bar{\mathbf{C}}_\nu^{+}
	\triangleq
	\bar{\mathbf{C}}_\nu
	+ \varepsilon_\nu(\mathbf{V}_\nu)\mathbf{I}_M.
	\label{eq:PB_C_upper}
\end{equation}
where
$\bar{\mathbf C}_{\nu}
=\mathbf G_{\nu}^{H}\mathbf V_{\nu}\mathbf G_{\nu}$. Accordingly, for any fixed positive semidefinite matrix $\mathbf Y\succeq\mathbf0$, the corresponding lower- and upper-bound terms can be expressed as
\begin{equation}
	\operatorname{Tr}\!\left(\bar{\mathbf{C}}_\nu^{-}\mathbf{Y}\right)
	=
	\operatorname{Tr}\!\left(
	\mathbf{G}_\nu
	\left[\mathbf{Y}
	-\hat{\varepsilon}\operatorname{Tr}(\mathbf{Y})\mathbf{I}_M\right]
	\mathbf{G}_\nu^H\mathbf{V}_\nu
	\right).
	\label{eq:PB_trace_lower}
\end{equation}
and
\begin{equation}
	\operatorname{Tr}\!\left(\bar{\mathbf{C}}_\nu^{+}\mathbf{Y}\right)
	=
	\operatorname{Tr}\!\left(
	\mathbf{G}_\nu
	\left[\mathbf{Y}
	+\hat{\varepsilon}\operatorname{Tr}(\mathbf{Y})\mathbf{I}_M\right]
	\mathbf{G}_\nu^H\mathbf{V}_\nu
	\right).
	\label{eq:PB_trace_upper}
\end{equation}

We next define the quantities associated with common-stream decoding as
\begin{equation}
	\begin{aligned}
		\Phi_{{\rm c},\nu}
		\left(
		\mathbf V_{\nu}
		\right)
		\triangleq\,
		&
		\varrho_{\nu}
		\operatorname{Tr}
		\left(
		\bar{\mathbf C}_{\nu}^{-}\mathbf X_{\rm c}
		\right)
		+
		\varrho_{\nu}
		\operatorname{Tr}
		\left(
		\bar{\mathbf C}_{\nu}^{+}\mathbf J_{{\rm c},\nu}
		\right)
		+
		\sigma_{{\rm ID},\nu}^{2},
	\end{aligned}
	\label{eq:PB_Phi_c}
\end{equation}
and
\begin{equation}
	\Psi_{{\rm c},\nu}
	\left(
	\mathbf V_{\nu}
	\right)
	\triangleq
	\varrho_{\nu}
	\operatorname{Tr}
	\left(
	\bar{\mathbf C}_{\nu}^{+}\mathbf J_{{\rm c},\nu}
	\right)
	+
	\sigma_{{\rm ID},\nu}^{2}.
	\label{eq:PB_Psi_c}
\end{equation}
Similarly, the corresponding quantities for private-stream decoding are defined as
\begin{equation}
	\begin{aligned}
		\Phi_{{\rm p},\nu}
		\left(
		\mathbf V_{\nu}
		\right)
		\triangleq\,
		&
		\varrho_{\nu}
		\operatorname{Tr}
		\left(
		\bar{\mathbf C}_{\nu}^{-}\mathbf X_{\nu}
		\right)
		+
		\varrho_{\nu}
		\operatorname{Tr}
		\left(
		\bar{\mathbf C}_{\nu}^{+}\mathbf J_{{\rm p},\nu}
		\right)
		+
		\sigma_{{\rm ID},\nu}^{2},
	\end{aligned}
	\label{eq:PB_Phi_p}
\end{equation}
and
\begin{equation}
	\Psi_{{\rm p},\nu}
	\left(
	\mathbf V_{\nu}
	\right)
	\triangleq
	\varrho_{\nu}
	\operatorname{Tr}
	\left(
	\bar{\mathbf C}_{\nu}^{+}\mathbf J_{{\rm p},\nu}
	\right)
	+
	\sigma_{{\rm ID},\nu}^{2}.
	\label{eq:PB_Psi_p}
\end{equation}

The robust rates for the common and private streams are consequently written as
\begin{equation}
	R_{{\rm c},\nu}^{\rm wc}
	\left(
	\mathbf V_{\nu}
	\right)
	=
	\log_{2}
	\Phi_{{\rm c},\nu}
	\left(
	\mathbf V_{\nu}
	\right)
	-
	\log_{2}
	\Psi_{{\rm c},\nu}
	\left(
	\mathbf V_{\nu}
	\right),
	\label{eq:PB_Rc}
\end{equation}
and
\begin{equation}
	R_{{\rm p},\nu}^{\rm wc}
	\left(
	\mathbf V_{\nu}
	\right)
	=
	\log_{2}
	\Phi_{{\rm p},\nu}
	\left(
	\mathbf V_{\nu}
	\right)
	-
	\log_{2}
	\Psi_{{\rm p},\nu}
	\left(
	\mathbf V_{\nu}
	\right).
	\label{eq:PB_Rp}
\end{equation}

The harvested-power requirement can be handled using the equivalent RF-input threshold $P_{{\rm RF},\nu}^{\rm req}$ derived in \eqref{eq:AB_RF_threshold}. Hence, the EH constraint is equivalently written as
\begin{equation}
	\begin{aligned}
		\scalemath{0.9}{(1-\varrho_{\nu})
		\Big[}
		&
		\scalemath{0.9}{\operatorname{Tr}
		\left(
		\bar{\mathbf C}_{\nu}^{-}
		\mathbf J_{{\rm EH},\nu}
		\right)
		+\sigma_{{\rm a},\nu}^{2}
		\Big]}
		&
		\geq
		P_{{\rm RF},\nu}^{\rm req}, 
		\qquad \forall\nu\in\mathcal S.
	\end{aligned}
	\label{eq:PB_EH_constraint}
\end{equation}
The left-hand side of \eqref{eq:PB_EH_constraint} is affine with respect to the passive beamforming matrices.

Although the above reformulation renders the received-power terms affine in $\mathbf V_{\nu}$, the rate expressions in \eqref{eq:PB_Rc} and \eqref{eq:PB_Rp} remain non-convex. At the $i$-th iteration, let $\mathbf V_{\nu}^{(i)}$ denote the corresponding expansion point and define $\Psi_{{\rm c},\nu}^{(i)}\triangleq\Psi_{{\rm c},\nu}(\mathbf V_{\nu}^{(i)})$ and $\Psi_{{\rm p},\nu}^{(i)}\triangleq\Psi_{{\rm p},\nu}(\mathbf V_{\nu}^{(i)})$. For compactness, define $\mathbf{B}_{s,\nu}\triangleq \mathbf{G}_\nu [\mathbf{J}_{s,\nu} +\hat{\varepsilon}\operatorname{Tr}(\mathbf{J}_{s,\nu})\mathbf{I}_M] \mathbf{G}_\nu^H$, where $s\in\{c,p\}$. By exploiting the concavity of $\log_2(x)$, a first-order upper-bounding surrogate is constructed around the current point. Accordingly,
\begin{align}
	\log_2\Psi_{c,\nu}(\mathbf{V}_\nu)
	\leq{}&
	\log_2\Psi_{c,\nu}^{(i)}
	+ \frac{
		\varrho_\nu\operatorname{Tr}\!\left[
		\mathbf{B}_{c,\nu}
		(\mathbf{V}_\nu-\mathbf{V}_\nu^{(i)})
		\right]
	}{\Psi_{c,\nu}^{(i)}\ln 2}
	\nonumber\\
	\triangleq{}&
	\widehat{\Psi}_{c,\nu}^{(i)}(\mathbf{V}_\nu),
	\quad \forall \nu\in\mathcal{S}.
	\label{eq:PB_Psi_c_upper}
\end{align}
and
\begin{align}
	\log_2\Psi_{p,\nu}(\mathbf{V}_\nu)
	\leq{}&
	\log_2\Psi_{p,\nu}^{(i)}
	+ \frac{
		\varrho_\nu\operatorname{Tr}\!\left[
		\mathbf{B}_{p,\nu}
		(\mathbf{V}_\nu-\mathbf{V}_\nu^{(i)})
		\right]
	}{\Psi_{p,\nu}^{(i)}\ln 2}
	\nonumber\\
	\triangleq{}&
	\widehat{\Psi}_{p,\nu}^{(i)}(\mathbf{V}_\nu),
	\quad \forall \nu\in\mathcal{S}.
	\label{eq:PB_Psi_p_upper}
\end{align}

Accordingly, the following expressions serve as lower bounds for the robust rates of the common and private streams:
\begin{equation}
	\widetilde R_{{\rm c},\nu}^{(i)}
	\left(
	\mathbf V_{\nu}
	\right)
	\triangleq
	\log_{2}
	\Phi_{{\rm c},\nu}
	\left(
	\mathbf V_{\nu}
	\right)
	-
	\widehat{\Psi}_{{\rm c},\nu}^{(i)}
	\left(
	\mathbf V_{\nu}
	\right),
	\label{eq:PB_Rc_lower}
\end{equation}
and
\begin{equation}
	\widetilde R_{{\rm p},\nu}^{(i)}
	\left(
	\mathbf V_{\nu}
	\right)
	\triangleq
	\log_{2}
	\Phi_{{\rm p},\nu}
	\left(
	\mathbf V_{\nu}
	\right)
	-
	\widehat{\Psi}_{{\rm p},\nu}^{(i)}
	\left(
	\mathbf V_{\nu}
	\right).
	\label{eq:PB_Rp_lower}
\end{equation}
These bounds satisfy
$\widetilde R_{{\rm c},\nu}^{(i)}
\leq R_{{\rm c},\nu}^{\rm wc}$ and
$\widetilde R_{{\rm p},\nu}^{(i)}
\leq R_{{\rm p},\nu}^{\rm wc}$,
with equality at
$\mathbf V_{\nu}=\mathbf V_{\nu}^{(i)}$.

Applying SDR to the rank-one constraints while retaining the trace-based uncertainty radius yields the following tractable relaxed reformulation:
\begin{subequations}
	\label{prob:P5}
	\begin{align}
		\mathrm{(P5)}\quad
		\max_{\mathcal Z_{\rm P}}\quad
		&
		\sum_{\nu\in\mathcal S}
		\left[
		\zeta_{\nu}
		+
		\widetilde R_{{\rm p},\nu}^{(i)}
		\left(
		\mathbf V_{\nu}
		\right)
		\right]
		\label{prob:P5a}
		\\
		\mathrm{s.t.}\quad
		&
		R_{\rm c}^{\rm tot}
		\leq
		\widetilde R_{{\rm c},\nu}^{(i)}
		\left(
		\mathbf V_{\nu}
		\right),
		\quad
		\forall\nu\in\mathcal S,
		\label{prob:P5b}
		\\
		&
		\zeta_{\nu}
		+
		\widetilde R_{{\rm p},\nu}^{(i)}
		\left(
		\mathbf V_{\nu}
		\right)
		\geq
		R_{\nu}^{\min},
		\quad
		\forall\nu\in\mathcal S,
		\label{prob:P5c}
		\\
		&
		\operatorname{diag}(\mathbf V_{\rm R})
		+
		\operatorname{diag}(\mathbf V_{\rm T})
		\leq
		\mathbf 1_{L},
		\label{prob:P5d}
		\\
		&
		\mathbf V_{\rm R}\succeq\mathbf0,
		\qquad
		\mathbf V_{\rm T}\succeq\mathbf0,
		\label{prob:P5e}
		\\
		&
		\Phi_{{\rm c},\nu}
		\left(
		\mathbf V_{\nu}
		\right)
		\geq
		\epsilon,
		\quad
		\Phi_{{\rm p},\nu}
		\left(
		\mathbf V_{\nu}
		\right)
		\geq
		\epsilon,
		\label{prob:P5f}
		\\
		&
		\eqref{prob:P1l}, \eqref{eq:PB_EH_constraint}.
		\label{prob:P5h}
	\end{align}
\end{subequations}
where
$\mathcal Z_{\rm P}
\triangleq
\left\{
\mathbf V_{\rm R},
\mathbf V_{\rm T},
\{\zeta_{\nu}\}_{\nu\in\mathcal S}
\right\}$,
and $\epsilon>0$ is a sufficiently small constant introduced to ensure that the logarithmic arguments remain strictly positive. Problem \eqref{prob:P5} is convex and can therefore be efficiently solved using standard convex optimization tools. After convergence, if $\mathbf V_R^\star$ and $\mathbf V_T^\star$ are rank one, the passive beamforming vectors $\boldsymbol{\theta}_R^\star$ and $\boldsymbol{\theta}_T^\star$ can be recovered via eigenvalue decomposition; otherwise, Gaussian randomization is employed to generate feasible candidate passive beamforming vectors \cite{ni2021resource}.

\subsection{ME-STARS Element-Position Optimization}
\label{subsec:position_optimization}
We next optimize the ME-STARS element positions by fixing the BS precoders, the reflection/transmission coefficients, and the PS ratios. The position matrix $\mathbf P$ and the common-rate allocation variables $\{\zeta_{\nu}\}_{\nu\in\mathcal S}$ are jointly updated. Using the equivalent RF-input threshold introduced in
\eqref{eq:AB_RF_threshold}, the corresponding position-design subproblem is formulated as
\begin{subequations}
	\label{prob:P6}
	\begin{align}
		\mathrm{(P6)}\quad
		\max_{\mathbf P,\{\zeta_{\nu}\}}\quad
		&
		\sum_{\nu\in\mathcal S}
		\left(
		\zeta_{\nu}
		+
		R_{{\rm p},\nu}^{\rm wc}(\mathbf P)
		\right)
		\label{prob:P6a}\\
		\mathrm{s.t.}\quad
		&
		R_{\rm c}^{\rm tot}
		\leq
		R_{{\rm c},\nu}^{\rm wc}(\mathbf P),
		\quad
		\forall\nu\in\mathcal S,
		\label{prob:P6b}\\
		&
		\zeta_{\nu}
		+
		R_{{\rm p},\nu}^{\rm wc}(\mathbf P)
		\geq
		R_{\nu}^{\min},
		\quad
		\forall\nu\in\mathcal S,
		\label{prob:P6c}
		\end{align}
		\begin{align}
		&
		P_{{\rm RF},\nu}^{\rm lb}(\mathbf P)
		\geq
		P_{{\rm RF},\nu}^{\rm req},
		\quad
		\forall\nu\in\mathcal S,
		\label{prob:P6d}\\
		& \eqref{prob:P1g}, \eqref{prob:P1h}, \eqref{prob:P1l}.
		\label{prob:P6e}
	\end{align}
\end{subequations}
Problem \eqref{prob:P6} is challenging mainly because the cascaded channels vary non-linearly with the movable-element positions, while the spacing requirement in \eqref{prob:P1h} is itself non-convex. We therefore
update the movable elements sequentially, while keeping the positions of the remaining elements unchanged.

For the subsequent derivation, we define
\begin{equation}
	\mathbf d(\theta,\psi)
	\triangleq
	\begin{bmatrix}
		\sin(\theta)\cos(\psi)\\
		\sin(\psi)
	\end{bmatrix},
	\qquad
	\mathbf v(\theta,\psi)
	\triangleq
	k_{\lambda}\mathbf d(\theta,\psi).
	\label{eq:PO_direction}
\end{equation}
Accordingly, $\mathbf v_{\rm A}\triangleq \mathbf v(\theta_{\rm A},\psi_{\rm A})$ denotes the spatial wave
vector associated with the BS--ME-STARS LoS component, whereas $\mathbf v_{\nu}\triangleq \mathbf v(\theta_{\nu},\psi_{\nu})$ denotes the corresponding spatial wave vector of the ME-STARS--user $\nu$ link. For notational convenience,
\begin{equation}
	(\theta_{\nu},\psi_{\nu},\gamma_{\nu})
	=
	\begin{cases}
		(\theta_{{\rm R},r},\psi_{{\rm R},r},
		\gamma_{{\rm R},r}),
		& \nu=({\rm R},r),\\[1mm]
		(\theta_{{\rm T},t},\psi_{{\rm T},t},
		\gamma_{{\rm T},t}),
		& \nu=({\rm T},t).
	\end{cases}
	\label{eq:PO_generic_parameters}
\end{equation}

Consider the update of the $\ell$-th movable element. At the $i$-th inner iteration, its current position is denoted by $\mathbf p_{\ell}^{(i)}$, while $\{\mathbf p_{\ell'}\}_{\ell'\neq\ell}$ remain fixed. For convenience, we use the column representation of the estimated cascaded channel, given by
\begin{equation}
	\mathbf h_{\nu}(\mathbf P)
	\triangleq
	\widehat{\mathbf z}_{\nu}^{H}(\mathbf P)
	=
	\mathbf F_{\rm B}^{H}(\mathbf P)
	\mathbf\Omega_{\nu}
	\mathbf f_{\nu}(\mathbf P).
	\label{eq:PO_h}
\end{equation}
For any fixed Hermitian matrix $\mathbf Y$, define the
position-dependent quadratic function as
\begin{equation}
	\mathcal G_{\nu,\mathbf Y}(\mathbf p_{\ell})
	\triangleq
	\mathbf h_{\nu}^{H}(\mathbf P)
	\mathbf Y
	\mathbf h_{\nu}(\mathbf P).
	\label{eq:PO_G}
\end{equation}

To incorporate the position-dependent uncertainty radius,
for $\mathbf Y\succeq\mathbf0$, define
$\mathbf{A}_{\mathbf{Y}}^\pm\triangleq
\mathbf{Y}\pm\hat{\varepsilon}
\operatorname{Tr}(\mathbf{Y})\mathbf{I}_M$.
The robust trace terms can then be expressed as
\begin{equation}
	\operatorname{Tr}\!\left(
	\mathbf{C}_\nu^-(\mathbf{P})\mathbf{Y}
	\right)
	=
	\mathcal G_{\nu,\mathbf{A}_{\mathbf{Y}}^-}
	(\mathbf p_\ell),
	\label{eq:PO_trace_lower}
\end{equation}
and
\begin{equation}
	\operatorname{Tr}\!\left(
	\mathbf{C}_\nu^+(\mathbf{P})\mathbf{Y}
	\right)
	=
	\mathcal G_{\nu,\mathbf{A}_{\mathbf{Y}}^+}
	(\mathbf p_\ell).
	\label{eq:PO_trace_upper}
\end{equation}

To obtain the derivatives of \eqref{eq:PO_G}, let $\mathbf e_{\ell}$ denote the $\ell$-th canonical basis vector of $\mathbb R^{L}$ and define
$\alpha_{\rm B}\triangleq\sqrt{\beta_{\rm B}\rho_{\rm B}/(\rho_{\rm B}+1)}$
and
$\eta_{\rm B}\triangleq\sqrt{\beta_{\rm B}/[(\rho_{\rm B}+1)R_{\rm B}]}$.
The position-dependent steering coefficients associated with element $\ell$ are defined as
$u_{{\rm A},\ell}\triangleq e^{j\mathbf v_{\rm A}^{T}\mathbf p_{\ell}}$,
$u_{r,\ell}\triangleq e^{j\mathbf v_r^{T}\mathbf p_{\ell}}$,
and
$f_{\nu,\ell}\triangleq\sqrt{\gamma_{\nu}}e^{j\mathbf v_{\nu}^{T}\mathbf p_{\ell}}$,
where
$\mathbf v_r\triangleq k_\lambda[\sin\theta_r\cos\psi_r,\sin\psi_r]^T$.
Accordingly, for $a,b\in\{x,y\}$, the first- and second-order derivatives of the BS--ME-STARS channel with respect to the coordinates of element $\ell$ are
\begin{equation}
	\begin{aligned}
		\mathbf F_{{\rm B},a}^{(\ell)}
		={}&
		j\alpha_{\rm B}v_{{\rm A},a}u_{{\rm A},\ell}
		\mathbf e_\ell\mathbf a_{\rm B}^{H}(\theta_{\rm B})
		\\
		&+
		j\eta_{\rm B}
		\sum_{r=1}^{R_{\rm B}}
		g_r v_{r,a}u_{r,\ell}
		\mathbf e_\ell\mathbf a_{\rm B}^{H}(\vartheta_r),
	\end{aligned}
	\label{eq:PO_FB_first}
\end{equation}
\begin{equation}
	\begin{aligned}
		\mathbf F_{{\rm B},ab}^{(\ell)}
		={}&
		-\alpha_{\rm B}v_{{\rm A},a}v_{{\rm A},b}
		u_{{\rm A},\ell}
		\mathbf e_\ell\mathbf a_{\rm B}^{H}(\theta_{\rm B})
		\\
		&-
		\eta_{\rm B}
		\sum_{r=1}^{R_{\rm B}}
		g_r v_{r,a}v_{r,b}u_{r,\ell}
		\mathbf e_\ell\mathbf a_{\rm B}^{H}(\vartheta_r),
	\end{aligned}
	\label{eq:PO_FB_second}
\end{equation}
respectively.

%\begin{equation}
%	\mathbf F_{{\rm B},a}^{(\ell)}
%	\triangleq
%	\frac{\partial\mathbf F_{\rm B}}
%	{\partial p_{a,\ell}}
%	=
%	j\alpha_{\rm B}
%	v_{{\rm A},a}
%	u_{{\rm A},\ell}
%	\mathbf e_{\ell}
%	\mathbf a_{\rm B}^{H}(\theta_{\rm B}),
%	\label{eq:PO_FB_first}
%\end{equation}

%\begin{equation}
%	\mathbf F_{{\rm B},ab}^{(\ell)}
%	\triangleq
%	\frac{\partial^{2}\mathbf F_{\rm B}}
%	{\partial p_{a,\ell}\partial p_{b,\ell}}
%	=
%	-\alpha_{\rm B}
%	v_{{\rm A},a}v_{{\rm A},b}
%	u_{{\rm A},\ell}
%	\mathbf e_{\ell}
%	\mathbf a_{\rm B}^{H}(\theta_{\rm B}).
%	\label{eq:PO_FB_second}
%\end{equation}

Similarly, the derivatives of the ME-STARS--user channel are
\begin{equation}
	\mathbf f_{\nu,a}^{(\ell)}
	\triangleq
	\frac{\partial\mathbf f_{\nu}}
	{\partial p_{a,\ell}}
	=
	jv_{\nu,a}
	f_{\nu,\ell}\mathbf e_{\ell},
	\label{eq:PO_f_first}
\end{equation}
and
\begin{equation}
	\mathbf f_{\nu,ab}^{(\ell)}
	\triangleq
	\frac{\partial^{2}\mathbf f_{\nu}}
	{\partial p_{a,\ell}\partial p_{b,\ell}}
	=
	-v_{\nu,a}v_{\nu,b}
	f_{\nu,\ell}\mathbf e_{\ell}.
	\label{eq:PO_f_second}
\end{equation}
Using \eqref{eq:PO_h}, the corresponding first-order derivative of the cascaded channel is
\begin{equation}
	\mathbf h_{\nu,a}^{(\ell)}
	\triangleq
	\frac{\partial\mathbf h_{\nu}}
	{\partial p_{a,\ell}}
	=
	\left(
	\mathbf F_{{\rm B},a}^{(\ell)}
	\right)^{H}
	\mathbf\Omega_{\nu}\mathbf f_{\nu}
	+
	\mathbf F_{\rm B}^{H}
	\mathbf\Omega_{\nu}
	\mathbf f_{\nu,a}^{(\ell)},
	\label{eq:PO_h_first}
\end{equation}
whereas its second-order derivative is

	\begin{align}
		\mathbf h_{\nu,ab}^{(\ell)}
		\triangleq\;&
		\frac{\partial^{2}\mathbf h_{\nu}}
		{\partial p_{a,\ell}\partial p_{b,\ell}}=
		\left(
		\mathbf F_{{\rm B},ab}^{(\ell)}
		\right)^{H}
		\mathbf\Omega_{\nu}\mathbf f_{\nu}
		+
		\left(
		\mathbf F_{{\rm B},a}^{(\ell)}
		\right)^{H}
		\mathbf\Omega_{\nu}
		\mathbf f_{\nu,b}^{(\ell)} \nonumber
		\\
		&+
		\left(
		\mathbf F_{{\rm B},b}^{(\ell)}
		\right)^{H}
		\mathbf\Omega_{\nu}
		\mathbf f_{\nu,a}^{(\ell)}
		+
		\mathbf F_{\rm B}^{H}
		\mathbf\Omega_{\nu}
		\mathbf f_{\nu,ab}^{(\ell)}.
	\end{align}\label{eq:PO_h_second}

The first-order partial derivative of
$\mathcal G_{\nu,\mathbf A_{\mathbf Y}^{\pm}}(\mathbf p_{\ell})$
is
\begin{equation}
	\frac{\partial
		\mathcal G_{\nu,\mathbf A_{\mathbf Y}^{\pm}}}
	{\partial p_{a,\ell}}
	=
	2\operatorname{Re}
	\left\{
	\left(
	\mathbf h_{\nu,a}^{(\ell)}
	\right)^{H}
	\mathbf A_{\mathbf Y}^{\pm}
	\mathbf h_{\nu}
	\right\},
	\quad a\in\{x,y\},
	\label{eq:PO_G_gradient}
\end{equation}
while its second-order partial derivatives are
\begin{equation}
	\begin{aligned}
		\frac{\partial^{2}
			\mathcal G_{\nu,\mathbf A_{\mathbf Y}^{\pm}}}
		{\partial p_{a,\ell}\partial p_{b,\ell}}
		=
		2\operatorname{Re}
		\Big\{
		&\scalemath{0.9}{
		\left(
		\mathbf h_{\nu,a}^{(\ell)}
		\right)^{H}
		\mathbf A_{\mathbf Y}^{\pm}
		\mathbf h_{\nu,b}^{(\ell)}
		+
		\mathbf h_{\nu}^{H}
		\mathbf A_{\mathbf Y}^{\pm}
		\mathbf h_{\nu,ab}^{(\ell)}
		\Big\}},
	\end{aligned}
	\label{eq:PO_G_Hessian}
\end{equation}
for $a,b\in\{x,y\}$. The corresponding gradient vector and Hessian matrix are therefore formed from the first- and second-order derivatives in \eqref{eq:PO_G_gradient} and \eqref{eq:PO_G_Hessian}, respectively. 

We next construct tractable quadratic bounds for the position-dependent functions $\mathcal G_{\nu,\mathbf A_{\mathbf Y}^{\pm}}(\mathbf p_{\ell})$ around the current position $\mathbf p_{\ell}^{(i)}$. For notational convenience, define the position displacement as
\begin{equation}
	\Delta\mathbf p_{\ell}
	\triangleq
	\mathbf p_{\ell}
	-
	\mathbf p_{\ell}^{(i)}.
	\label{eq:PO_displacement}
\end{equation}
The following lemma provides the required lower and upper bounds.

\noindent\textbf{Lemma 1:}
For the twice continuously differentiable function $\mathcal G_{\nu,\mathbf A_{\mathbf Y}^{\pm}}(\mathbf p_{\ell})$, suppose that a curvature parameter $\chi_{\nu,\mathbf A_{\mathbf Y}^{\pm}}^{(i)}\geq0$ can be chosen such that
\begin{equation}
	-\chi_{\nu,\mathbf A_{\mathbf Y}^{\pm}}^{(i)}\mathbf I_{2}
	\preceq
	\nabla^{2}
	\mathcal G_{\nu,\mathbf A_{\mathbf Y}^{\pm}}(\mathbf p_{\ell})
	\preceq
	\chi_{\nu,\mathbf A_{\mathbf Y}^{\pm}}^{(i)}\mathbf I_{2},
	\quad
	\forall\mathbf p_{\ell}\in\mathcal A.
	\label{eq:PO_curvature}
\end{equation}
Then, a concave quadratic lower bound of $\mathcal G_{\nu,\mathbf A_{\mathbf Y}^{\pm}}(\mathbf p_{\ell})$ around $\mathbf p_{\ell}^{(i)}$ is given by
\begin{align}
	\underline{\mathcal G}_{\nu,\mathbf A_{\mathbf Y}^{\pm}}^{(i)}
	(\mathbf p_{\ell})
	\triangleq\;&
	\mathcal G_{\nu,\mathbf A_{\mathbf Y}^{\pm}}
	(\mathbf p_{\ell}^{(i)})+
\nabla\mathcal G_{\nu,\mathbf A_{\mathbf Y}^{\pm}}
	(\mathbf p_{\ell}^{(i)})^{T}
	\Delta\mathbf p_{\ell}
	\nonumber\\
	&-
	\frac{\chi_{\nu,\mathbf A_{\mathbf Y}^{\pm}}^{(i)}}{2}
	\|\Delta\mathbf p_{\ell}\|_{2}^{2},
	\label{eq:PO_G_lower}
\end{align}
whereas a convex quadratic upper bound is
\begin{align}
	\overline{\mathcal G}_{\nu,\mathbf A_{\mathbf Y}^{\pm}}^{(i)}
	(\mathbf p_{\ell})
	\triangleq\;&
	\mathcal G_{\nu,\mathbf A_{\mathbf Y}^{\pm}}
	(\mathbf p_{\ell}^{(i)})+
	\nabla\mathcal G_{\nu,\mathbf A_{\mathbf Y}^{\pm}}
	(\mathbf p_{\ell}^{(i)})^{T}
	\Delta\mathbf p_{\ell}
	\nonumber\\
	&+
	\frac{\chi_{\nu,\mathbf A_{\mathbf Y}^{\pm}}^{(i)}}{2}
	\|\Delta\mathbf p_{\ell}\|_{2}^{2}.
	\label{eq:PO_G_upper}
\end{align}
Accordingly,
\begin{equation}
	\underline{\mathcal G}_{\nu,\mathbf A_{\mathbf Y}^{\pm}}^{(i)}
	(\mathbf p_{\ell})
	\leq
	\mathcal G_{\nu,\mathbf A_{\mathbf Y}^{\pm}}(\mathbf p_{\ell})
	\leq
	\overline{\mathcal G}_{\nu,\mathbf A_{\mathbf Y}^{\pm}}^{(i)}
	(\mathbf p_{\ell}),
	\quad
	\forall\mathbf p_{\ell}\in\mathcal A,
	\label{eq:PO_G_bounds}
\end{equation}
where both bounds are tight and gradient-consistent with $\mathcal G_{\nu,\mathbf A_{\mathbf Y}^{\pm}}(\mathbf p_{\ell})$ at $\mathbf p_{\ell}=\mathbf p_{\ell}^{(i)}$. The lemma 1 guarantees valid quadratic bounds over $\mathcal A$ when its curvature condition holds. In practice, the curvature parameter is initialized using the current Hessian norm and increased until all quadratic bounds hold at the candidate position, re-solving the subproblem after each increase. These checks ensure feasibility and a nondecreasing objective at accepted updates. 

Using \eqref{eq:PO_trace_lower} and \eqref{eq:PO_trace_upper}, the common-stream rate can be written as
\begin{equation}
	R_{{\rm c},\nu}^{\rm wc}(\mathbf p_{\ell})
	=
	\log_{2}\mathcal N_{{\rm c},\nu}(\mathbf p_{\ell})
	-
	\log_{2}\mathcal D_{{\rm c},\nu}(\mathbf p_{\ell}),
	\label{eq:PO_Rc}
\end{equation}
where
\begin{align}
	\mathcal N_{{\rm c},\nu}(\mathbf p_{\ell})
	\triangleq\;&
	\varrho_{\nu}
	\mathcal G_{\nu,\mathbf A_{\mathbf X_{\rm c}}^{-}}
	(\mathbf p_{\ell})+
	\varrho_{\nu}
	\mathcal G_{\nu,\mathbf A_{\mathbf J_{{\rm c},\nu}}^{+}}
	(\mathbf p_{\ell})
	+\sigma_{{\rm ID},\nu}^{2},
	\label{eq:PO_Nc}
\end{align}
\begin{equation}
	\mathcal D_{{\rm c},\nu}(\mathbf p_{\ell})
	\triangleq
	\varrho_{\nu}
	\mathcal G_{\nu,\mathbf A_{\mathbf J_{{\rm c},\nu}}^{+}}
	(\mathbf p_{\ell})
	+\sigma_{{\rm ID},\nu}^{2}.
	\label{eq:PO_Dc}
\end{equation}
Likewise, the private-stream rate is expressed as
\begin{equation}
	R_{{\rm p},\nu}^{\rm wc}(\mathbf p_{\ell})
	=
	\log_{2}\mathcal N_{{\rm p},\nu}(\mathbf p_{\ell})
	-
	\log_{2}\mathcal D_{{\rm p},\nu}(\mathbf p_{\ell}),
	\label{eq:PO_Rp}
\end{equation}
where
\begin{align}
	\mathcal N_{{\rm p},\nu}(\mathbf p_{\ell})
	\triangleq\;&
	\scalemath{0.95}{\varrho_{\nu}
	\mathcal G_{\nu,\mathbf A_{\mathbf X_{\nu}}^{-}}
	(\mathbf p_{\ell})+
	\varrho_{\nu}
	\mathcal G_{\nu,\mathbf A_{\mathbf J_{{\rm p},\nu}}^{+}}
	(\mathbf p_{\ell})
	+\sigma_{{\rm ID},\nu}^{2}},
	\label{eq:PO_Np}
\end{align}
\begin{equation}
	\mathcal D_{{\rm p},\nu}(\mathbf p_{\ell})
	\triangleq
	\varrho_{\nu}
	\mathcal G_{\nu,\mathbf A_{\mathbf J_{{\rm p},\nu}}^{+}}
	(\mathbf p_{\ell})
	+\sigma_{{\rm ID},\nu}^{2}.
	\label{eq:PO_Dp}
\end{equation}

For compactness, let $\mathbf X_s=\mathbf X_{\rm c}$ for
$s={\rm c}$ and $\mathbf X_s=\mathbf X_\nu$ for $s={\rm p}$.
Using \eqref{eq:PO_G_lower} and \eqref{eq:PO_G_upper},
concave lower approximations of the numerators and convex
upper approximations of the denominators are given by
\begin{align}
	\underline{\mathcal N}_{s,\nu}^{(i)}(\mathbf p_{\ell})
	\triangleq\;&
	\scalemath{0.95}{\varrho_{\nu}
	\underline{\mathcal G}_{\nu,\mathbf A_{\mathbf X_s}^{-}}^{(i)}
	(\mathbf p_{\ell})+
	\varrho_{\nu}
	\underline{\mathcal G}_{\nu,\mathbf A_{\mathbf J_{s,\nu}}^{+}}^{(i)}
	(\mathbf p_{\ell})
	+\sigma_{{\rm ID},\nu}^{2}},
	\label{eq:PO_N_lower}
\end{align}
\begin{equation}
	\overline{\mathcal D}_{s,\nu}^{(i)}(\mathbf p_{\ell})
	\triangleq
	\varrho_{\nu}
	\overline{\mathcal G}_{\nu,\mathbf A_{\mathbf J_{s,\nu}}^{+}}^{(i)}
	(\mathbf p_{\ell})
	+\sigma_{{\rm ID},\nu}^{2},
	\label{eq:PO_D_upper}
\end{equation}
for $s\in\{{\rm c},{\rm p}\}$.
Define
$\mathcal D_{s,\nu}^{(i)}
\triangleq\mathcal D_{s,\nu}(\mathbf p_{\ell}^{(i)})$.
By the concavity of $\log_2(x)$, a concave lower bound
on the corresponding robust rate is
\begin{align}
	\widetilde R_{s,\nu}^{(i)}(\mathbf p_{\ell})
	\triangleq\;&
	\log_{2}\underline{\mathcal N}_{s,\nu}^{(i)}
	(\mathbf p_{\ell})
	-\log_{2}\mathcal D_{s,\nu}^{(i)}
	\nonumber\\
	&-
	\frac{
		\overline{\mathcal D}_{s,\nu}^{(i)}(\mathbf p_{\ell})
		-\mathcal D_{s,\nu}^{(i)}
	}{
		\mathcal D_{s,\nu}^{(i)}\ln2
	},
	\label{eq:PO_rate_lower}
\end{align}
for $s\in\{{\rm c},{\rm p}\}$.
The bound in \eqref{eq:PO_rate_lower} satisfies
$\widetilde R_{s,\nu}^{(i)}(\mathbf p_{\ell})
\leq R_{s,\nu}^{\rm wc}(\mathbf p_{\ell})$
and is tight and gradient-consistent at
$\mathbf p_{\ell}=\mathbf p_{\ell}^{(i)}$.

The EH constraint can be handled using the same lower quadratic model. In particular, a concave lower bound on the RF power delivered to user
$\nu$ is
\begin{equation}
	\begin{aligned}
		\underline P_{{\rm RF},\nu}^{(i)}(\mathbf p_\ell)
		\triangleq
		(1-\varrho_\nu)\Big[
		&\underline{\mathcal G}_{
			\nu,\mathbf A_{\mathbf J_{{\rm EH},\nu}}^{-}}^{(i)}
		(\mathbf p_\ell)+\sigma_{{\rm a},\nu}^{2}
		\Big].
	\end{aligned}
		\label{eq:PO_EH_lower}
\end{equation}

It remains to handle the minimum inter-element spacing constraint. For each $\ell'\neq\ell$, define
\begin{equation}
	\mathbf b_{\ell,\ell'}^{(i)}
	\triangleq
	\frac{
		\mathbf p_{\ell}^{(i)}-\mathbf p_{\ell'}
	}{
		\|
		\mathbf p_{\ell}^{(i)}-\mathbf p_{\ell'}
		\|_{2}
	}.
	\label{eq:PO_spacing_direction}
\end{equation}
Since $\|\mathbf p_{\ell}-\mathbf p_{\ell'}\|_{2}$ is convex, its first-order approximation at $\mathbf p_{\ell}^{(i)}$ provides the global affine lower bound
\begin{equation}
	\|\mathbf p_{\ell}-\mathbf p_{\ell'}\|_{2}
	\geq
	\left(
	\mathbf b_{\ell,\ell'}^{(i)}
	\right)^{T}
	(\mathbf p_{\ell}-\mathbf p_{\ell'}).
	\label{eq:PO_spacing_lower}
\end{equation}
Hence, the minimum-spacing requirement can be conservatively replaced
by
\begin{equation}
	\left(
	\mathbf b_{\ell,\ell'}^{(i)}
	\right)^{T}
	(\mathbf p_{\ell}-\mathbf p_{\ell'})
	\geq
	d_{\rm s},
	\qquad
	\forall\ell'\neq\ell.
	\label{eq:PO_spacing_affine}
\end{equation}

At the $i$-th inner iteration, $\mathbf p_\ell$ is optimized jointly with $\{\zeta_\nu\}_{\nu\in\mathcal S}$ through the following problem:
\begin{subequations}
	\label{prob:P7}
	\begin{align}
		\mathrm{(P7)}\quad
		\max_{\mathcal Z_{\ell}}\quad
		&
		\sum_{\nu\in\mathcal S}
		\left[
		\zeta_{\nu}
		+
		\widetilde R_{{\rm p},\nu}^{(i)}
		(\mathbf p_{\ell})
		\right]
		\label{prob:P7a}\\
		\mathrm{s.t.}\quad
		&
		R_{\rm c}^{\rm tot}
		\leq
		\widetilde R_{{\rm c},\nu}^{(i)}
		(\mathbf p_{\ell}),
		\quad
		\forall\nu\in\mathcal S,
		\label{prob:P7b}\\
		&
		\zeta_{\nu}
		+
		\widetilde R_{{\rm p},\nu}^{(i)}
		(\mathbf p_{\ell})
		\geq
		R_{\nu}^{\min},
		\quad
		\forall\nu\in\mathcal S,
		\label{prob:P7c}\\
		&
		\underline P_{{\rm RF},\nu}^{(i)}
		(\mathbf p_{\ell})
		\geq
		P_{{\rm RF},\nu}^{\rm req},
		\quad
		\forall\nu\in\mathcal S,
		\label{prob:P7d}\\
		&
		\underline{\mathcal N}_{{\rm c},\nu}^{(i)}
		(\mathbf p_{\ell})
		\geq
		\epsilon,
		\quad
		\forall\nu\in\mathcal S,
		\label{prob:P7e}\\
		&
		\underline{\mathcal N}_{{\rm p},\nu}^{(i)}
		(\mathbf p_{\ell})
		\geq
		\epsilon,
		\quad
		\forall\nu\in\mathcal S,
		\label{prob:P7f}\\
		& \eqref{prob:P6e}, \eqref{eq:PO_spacing_affine}.
		\label{prob:P7g}
	\end{align}
\end{subequations}
where $\mathcal Z_{\ell} \triangleq \left\{\mathbf p_{\ell},\{\zeta_{\nu}\}_{\nu\in\mathcal S} \right\}$. Here, $\epsilon>0$ is a sufficiently small constant introduced to ensure positivity of the logarithmic arguments. Problem \eqref{prob:P7} is convex and can therefore be efficiently solved using standard convex optimization solvers. After convergence of the $\ell$-th element update, the optimization continues with the subsequent element. After all elements have been updated sequentially for $\ell=1,\ldots,L$, the resulting position matrix is used in the next iteration of the overall optimization algorithm.

\subsection{Power-Splitting Ratio Optimization}
\label{subsec:PS_optimization}
We finally optimize the PS ratios by fixing the BS precoders, ME-STARS passive beamforming, and spatial positions of the movable elements.
\begin{subequations}
	\label{prob:P8}
	\begin{align}
		\mathrm{(P8)}\quad
		\max_{\mathcal Z_{\rm S}}\quad
		&
		\sum_{\nu\in\mathcal S}
		\left(
		\zeta_{\nu}
		+
		R_{{\rm p},\nu}^{\rm wc}(\varrho_{\nu})
		\right)
		\label{prob:P8a}\\
		\mathrm{s.t.}\quad
		&
		R_{\rm c}^{\rm tot}
		\leq
		R_{{\rm c},\nu}^{\rm wc}(\varrho_{\nu}),
		\quad
		\forall\nu\in\mathcal S,
		\label{prob:P8b}\\
		&
		\zeta_{\nu}
		+
		R_{{\rm p},\nu}^{\rm wc}(\varrho_{\nu})
		\geq
		R_{\nu}^{\min},
		\quad
		\forall\nu\in\mathcal S,
		\label{prob:P8c}\\
		&
		\eqref{prob:P1f}, \eqref{prob:P1l},\eqref{eq:AB_EH_RF_constraint}.\label{prob:P8d}
	\end{align}
\end{subequations}
where $\mathcal Z_{\rm S} \triangleq \left\{\{\varrho_{\nu}\}_{\nu\in\mathcal S}, \{\zeta_{\nu}\}_{\nu\in\mathcal S}\right\}$. For fixed beamforming variables and element positions, $R_{{\rm c},\nu}^{\rm wc}(\varrho_{\nu})$ and $R_{{\rm p},\nu}^{\rm wc}(\varrho_{\nu})$ are concave functions of $\varrho_{\nu}$, while the EH constraint in \eqref{eq:AB_EH_RF_constraint} is affine. Hence, problem \eqref{prob:P8} is convex and can be efficiently solved using standard convex optimization solvers.

\subsection{Complexity Analysis of the Proposed Algorithm}
\label{subsec:complexity}
The proposed algorithm consists of four iterative blocks associated with active beamforming, ME-STARS passive beamforming, element-position optimization, and PS-ratio design. Let $I_{\rm A}$, $I_{\rm P}$, $I_{\rm Q}$, and $I_{\rm S}$ represent their respective iteration counts, while $I_{\rm O}$ denotes the number of outer-loop updates until convergence. Considering first the active-beamforming problem in \eqref{prob:P3}, one common-stream covariance matrix and $J$ private-stream covariance matrices of size $M\times M$ are optimized, resulting in a dominant complexity of $\mathcal O\!\left(I_{\rm A}(J+1)^{3.5}M^{7} \log(1/\epsilon_{\rm A})\right)$ \cite{luo2010semidefinite}. The passive beamforming update in \eqref{prob:P5} involves two $L\times L$ positive semidefinite matrices and has a complexity of $\mathcal O\!\left(I_{\rm P}L^{7} \log(1/\epsilon_{\rm P})\right)$ \cite{luo2010semidefinite}. For the movable-element position update, problem \eqref{prob:P7} is solved successively for the $L$ elements, with the channel derivatives and quadratic bounds evaluated for all $J$ users. The corresponding computational cost is $\mathcal O\!\left(I_{\rm Q}JL^{2} \log(1/\epsilon_{\rm Q})\right)$ \cite{grant2008cvx}. The PS-ratio subproblem in \eqref{prob:P8} only involves scalar optimization variables associated with the $J$ users, resulting in a comparatively lower complexity of $\mathcal O\!\left(I_{\rm S}J^{3}\right)$. Accordingly, the overall computational complexity of the proposed algorithm is $\mathcal O\bigg(I_{\rm O}\big[ I_{\rm A}(J+1)^{3.5}M^{7}\log(1/\epsilon_{\rm A}) +I_{\rm P}L^{7}\log(1/\epsilon_{\rm P}) +I_{\rm Q}JL^{2}$\\ $\log(1/\epsilon_{\rm Q}) +I_{\rm S}J^{3}\big]\bigg)$.

\begin{algorithm}[t]
	\caption{Robust Transmission Design for ME-STARS-Assisted RSMA-SWIPT}
	\label{alg:overall_optimization}
	\begin{algorithmic}[1]
		
		\State \textbf{Input:} Convergence tolerance $\epsilon_{\rm AO}$ and
		maximum number of outer iterations ${\mathcal I}_{\rm AO}^{\max}$.
		
		\State Generate a feasible initial point $\mathcal V^{(0)}$,
		compute its robust sum rate $R_{\rm sum}^{(0)}$, and set $t=0$.
		
		\Repeat
		
		\Statex \hspace{\algorithmicindent}\textit{Active-precoder design:}
		
		\State With $\mathbf\Omega_{\rm R}$, $\mathbf\Omega_{\rm T}$,
		$\mathbf P$, and $\{\varrho_{\nu}\}_{\nu\in\mathcal S}$ fixed,
		solve \eqref{prob:P3}.
		
		\State Obtain $\mathbf X_{\rm c}$,
		$\{\mathbf X_{\nu}\}_{\nu\in\mathcal S}$, and
		$\{\zeta_{\nu}\}_{\nu\in\mathcal S}$, and recover the corresponding
		BS precoders.
		
		\Statex \hspace{\algorithmicindent}\textit{ME-STARS passive-beamforming design:}
		
		\State For the updated BS precoders and fixed $\mathbf P$ and $\{\varrho_{\nu}\}_{\nu\in\mathcal S}$, solve \eqref{prob:P5}.
		
		\State Obtain $\mathbf V_{\rm R}$ and $\mathbf V_{\rm T}$,
		recover $\boldsymbol{\theta}_{\rm R}$ and
		$\boldsymbol{\theta}_{\rm T}$, and construct
		$\mathbf\Omega_{\rm R}$ and $\mathbf\Omega_{\rm T}$.
		
		\Statex \hspace{\algorithmicindent}\textit{Movable-element positioning:}
		
		\For{$\ell=1,\ldots,L$}
		
		\State Keep $\{\mathbf p_{\ell'}\}_{\ell'\neq\ell}$ fixed and
		initialize the inner counter $i=0$.
		
		\Repeat
		
		\State Evaluate the channel derivatives and curvature parameters
		associated with $\mathbf p_{\ell}^{(i)}$.
		
		\State Construct the surrogates for the rate, EH, and
		inter-element spacing terms.
		
		\State Solve \eqref{prob:P7} to obtain
		$\mathbf p_{\ell}^{\star}$ and
		$\{\zeta_{\nu}^{\star}\}_{\nu\in\mathcal S}$.
		
       \State Update $\mathbf p_{\ell}$ and
       $\{\zeta_{\nu}\}_{\nu\in\mathcal S}$ via backtracking,
       and set $i\leftarrow i+1$.
		
		\Until{the inner update converges or
			$i={\mathcal I}_{\rm Q}^{\max}$}
		
		Insert the optimized $\mathbf p_\ell$ into $\mathbf P$
		and refresh the channels and uncertainty radii.
		
		\EndFor
		
		\Statex \hspace{\algorithmicindent}\textit{PS-ratio design:}
		
		\State With the other variable blocks fixed, solve \eqref{prob:P8}
		and obtain $\{\varrho_{\nu}\}_{\nu\in\mathcal S}$ together with
		$\{\zeta_{\nu}\}_{\nu\in\mathcal S}$.
		
		\State Set $t\leftarrow t+1$.
		\State Compute the updated robust sum rate $R_{\rm sum}^{(t)}$.
		
		\Until{
			$\left|
			R_{\rm sum}^{(t)}
			-
			R_{\rm sum}^{(t-1)}
			\right|
			\leq\epsilon_{\rm AO}$
			or
			$t={\mathcal I}_{\rm AO}^{\max}$.}
		
		\State \Return $\mathcal V^{\star}$.
		
	\end{algorithmic}
\end{algorithm}
%%%%%%%%%%%%%%%%%%%%%%%%%%%%%%%%%%%%%%%%%%%%%%%%%%%%%%%%%%%%%%%%%%%%
\section{Results and Performance Analysis}
The performance of the proposed framework is evaluated for a downlink ME-STARS-assisted RSMA-SWIPT network with users distributed across the reflection and transmission regions. Unless otherwise stated, we set $M=8$, $L=16$, and $J=4$, with two users in each region. The wavelength is $\lambda=0.03$~m, the movement region is $3\lambda\times3\lambda$, and $d_s=\lambda/2$. The BS--ME-STARS distance is $20$~m, and the surface--user distances are $5$~m and $3$~m in each region. The BS--ME-STARS LoS angles are $\theta_{\rm B}=120^\circ$, $\theta_{\rm A}=330^\circ$, and $\psi_{\rm A}=30^\circ$. The user angles are $\boldsymbol{\theta}_{\rm T}=[140^\circ,210^\circ]$, $\boldsymbol{\psi}_{\rm T}=[-45^\circ,-30^\circ]$, $\boldsymbol{\theta}_{\rm R}=[-45^\circ,30^\circ]$, and $\boldsymbol{\psi}_{\rm R}=[-25^\circ,-10^\circ]$. For the BS--ME-STARS channel, we set $\rho_{\rm B}=3$ and $R_{\rm B}=3$. The NLoS parameters are independently generated as $\theta_r\sim\mathcal U[0,2\pi)$, $\psi_r\sim\mathcal U[-\pi/2,\pi/2]$, $\vartheta_r\sim\mathcal U[-\pi/2,\pi/2]$ \cite{geng2025joint}, and $g_r\sim\mathcal{CN}(0,1)$. These parameters remain fixed within each realization, while the steering vectors vary with the element positions. The ME-STARS--user channels are single-path LoS. The path gain is $\rho_0d^{-2}$, with $\rho_0=-30$~dB at $1$~m, and results are averaged over $10^3$ channel realizations. We set $P_{\rm tx}^{\max}=50$~dBm, $R_\nu^{\min}=1$~bps/Hz \cite{asif2026robust123}, $P_{{\rm EH},\nu}^{\min}=-35$~dBm, $\sigma_{{\rm a},\nu}^{2}=\sigma_{{\rm d},\nu}^{2}=-90$~dBm, and $\kappa_{\rm tx}=\kappa_{{\rm rx},\nu}=0.02$. The PS ratios are initialized at $\varrho_\nu=0.5$ and subsequently optimized. The nonlinear EH parameters are $P_\nu^{\rm sat}=0.2\,\mathrm{mW}$, $b_\nu=0.003$, and $a_\nu=6400$ {\cite{xiong2017rate}, $\forall \nu\in\mathcal S$. The relative CSI uncertainty is controlled by $\hat{\varepsilon}\in[0,1)$, with $\varepsilon_\nu$ evaluated as defined in Section~II-C \cite{asif2026robust,yao2023robust}.

\begin{figure}[!t]
	\centering
	\includegraphics [width=0.35\textwidth]{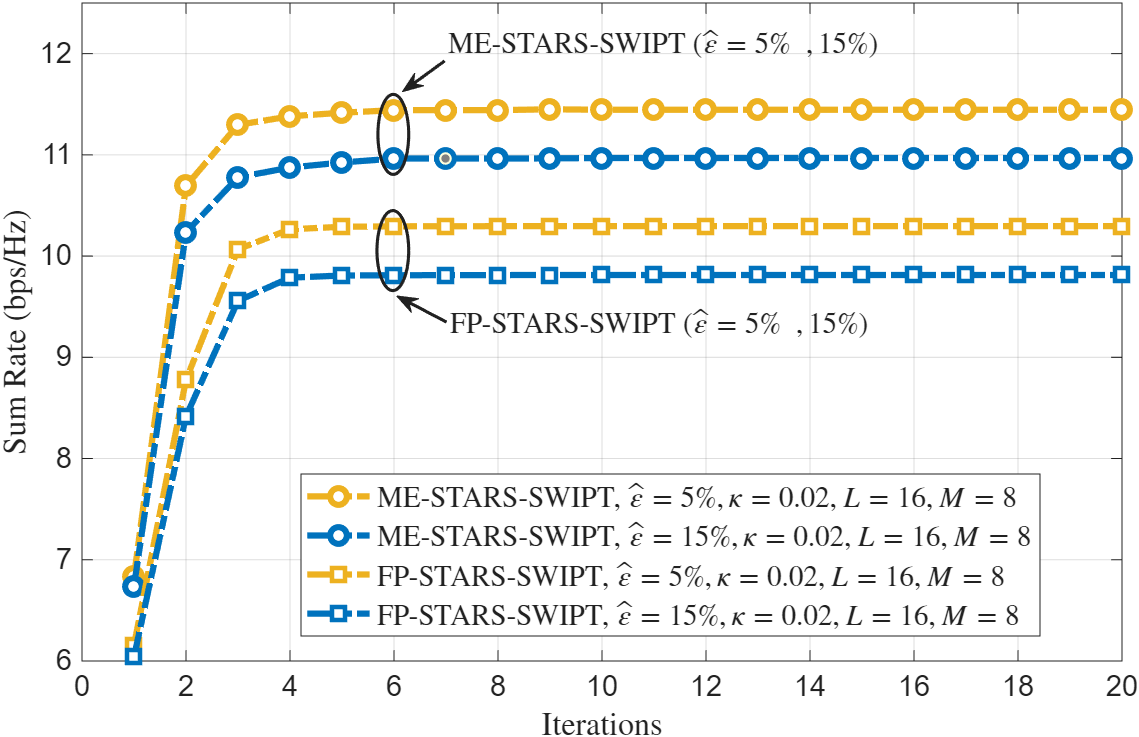}
	\caption{Convergence behavior under varying CSI uncertainty.}
	\label{f2}
\end{figure} 

\begin{figure}[t]
	\centering
	\includegraphics [width=0.35\textwidth]{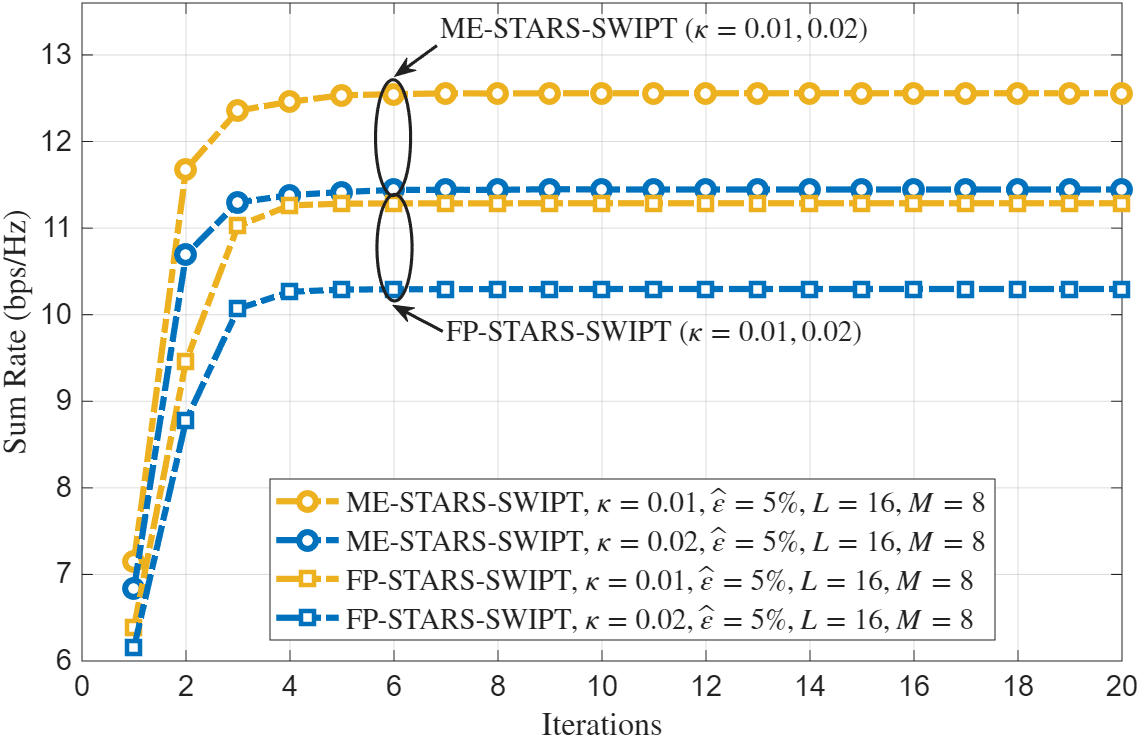}
	\caption{Convergence under different hardware-impairment levels.}
	\label{f3}
\end{figure}

\begin{figure}[t]
	\centering
	\includegraphics [width=0.35\textwidth]{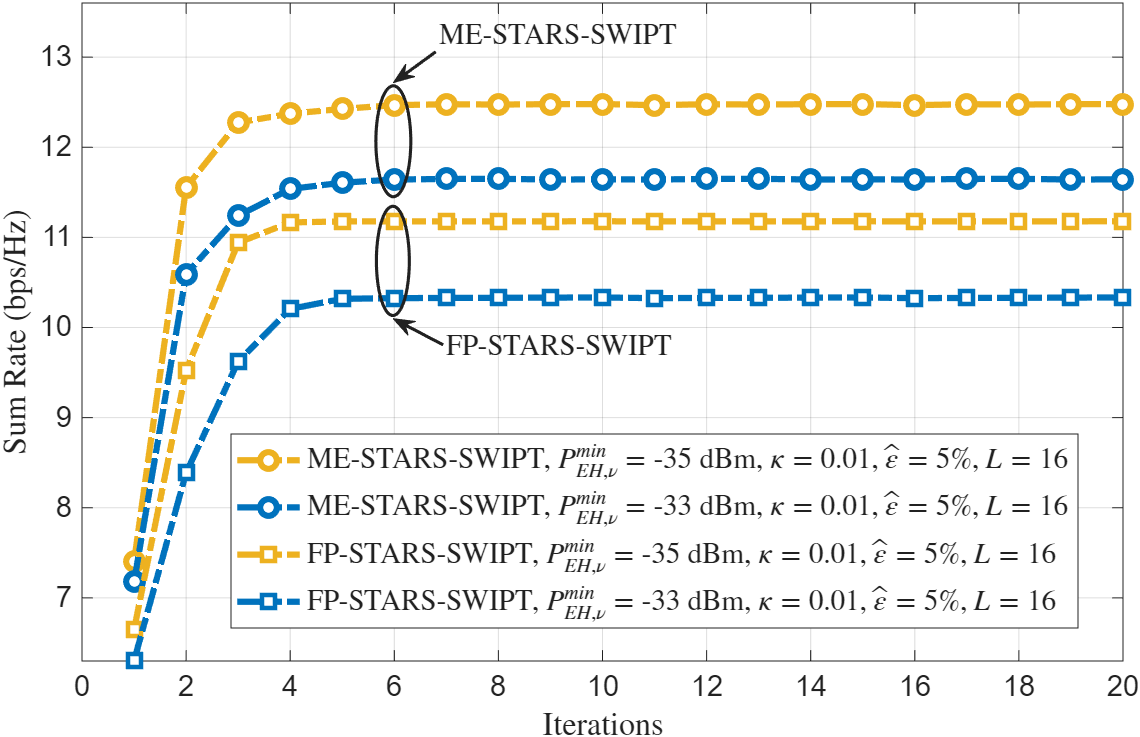}
	\caption{Convergence under different EH requirements.}
	\label{f4}
\end{figure}

  \begin{figure}[!h]
	\centering
	\includegraphics [width=0.35\textwidth]{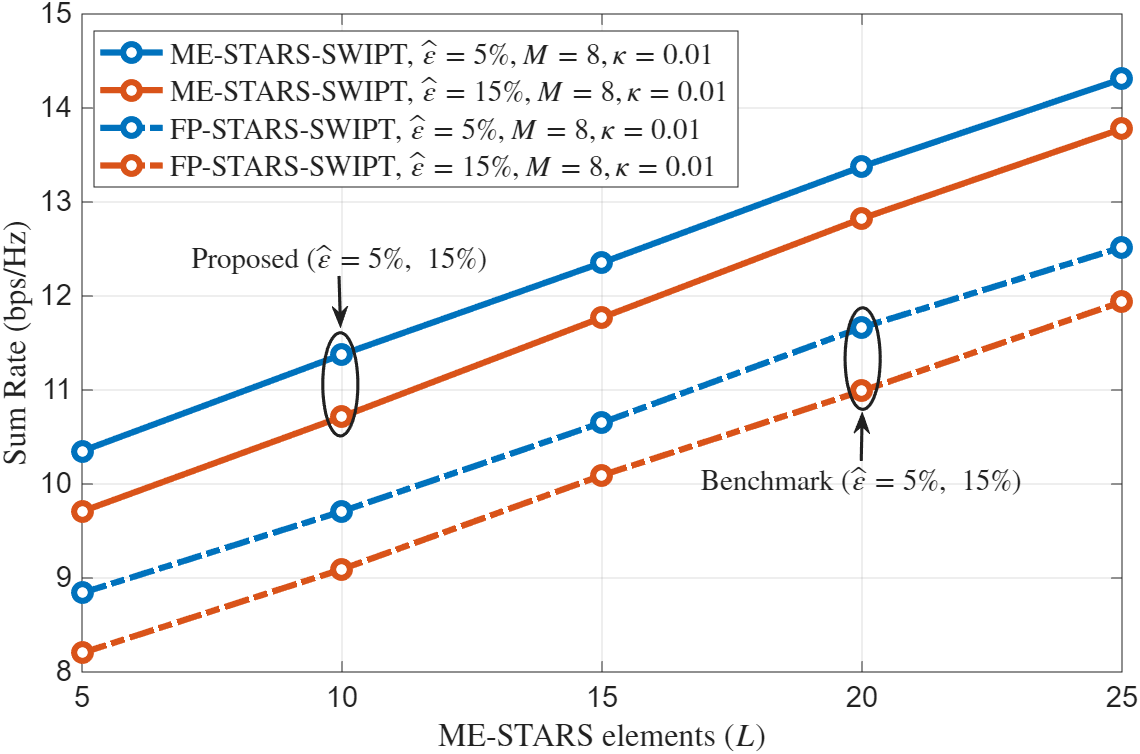}
	\caption{Sum rate versus $L$ under different CSI uncertainty levels.}
	\label{f5}
\end{figure}

\begin{figure}[h]
	\centering
	\includegraphics [width=0.35\textwidth]{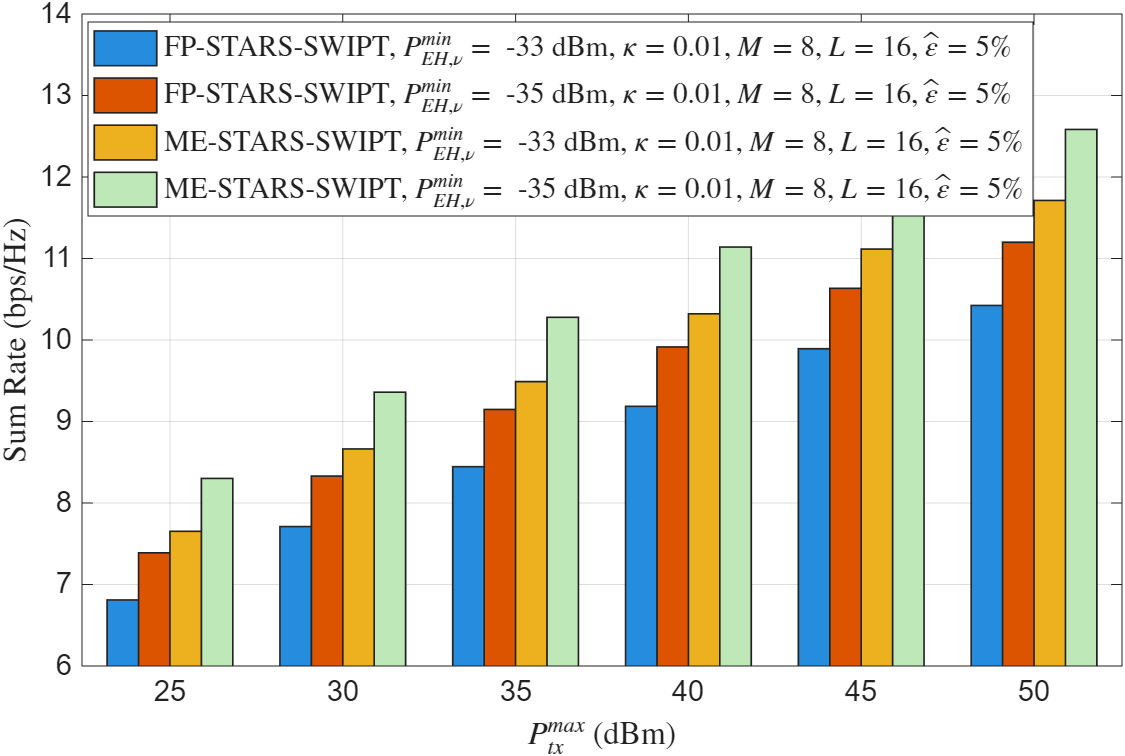}
	\caption{Sum rate versus $P_{\rm tx}^{\max}$ under different EH requirements.}
	\label{f6}
\end{figure}

\begin{figure}[h]
\centering
\includegraphics [width=0.35\textwidth]{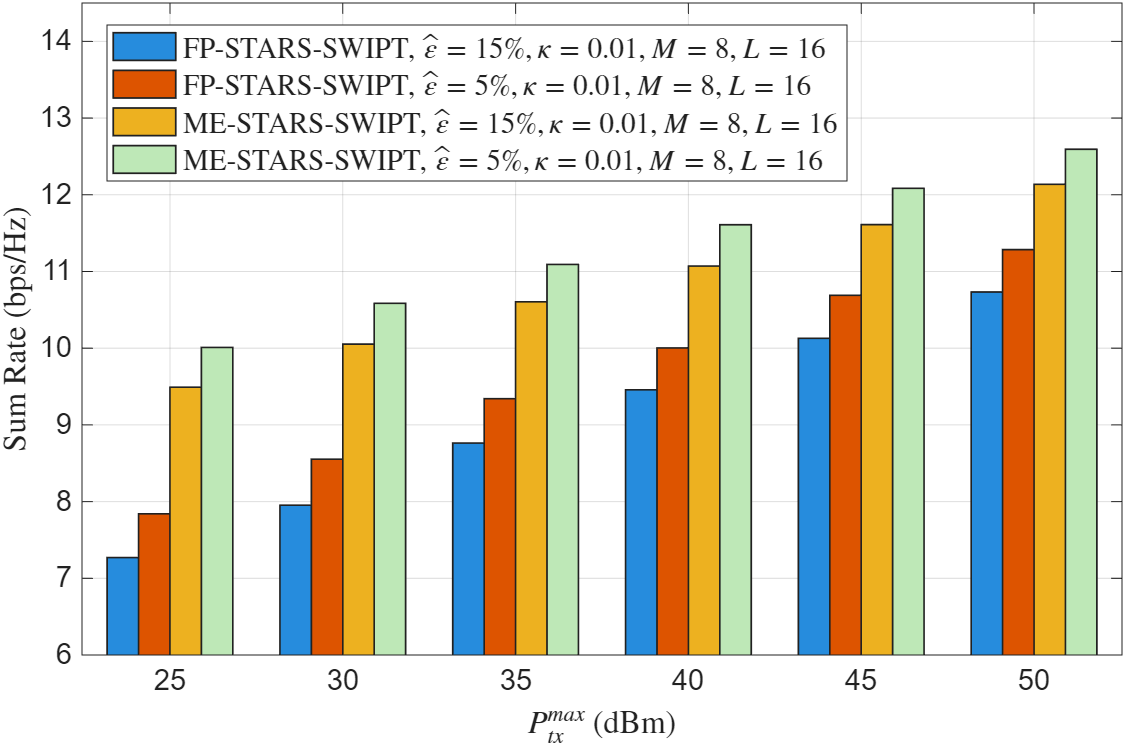}
\caption{Sum rate versus $P_{\rm tx}^{\max}$ under different CSI uncertainty levels.}
\label{f7}
\end{figure}

\begin{figure}[h]
	\centering
	\includegraphics [width=0.35\textwidth]{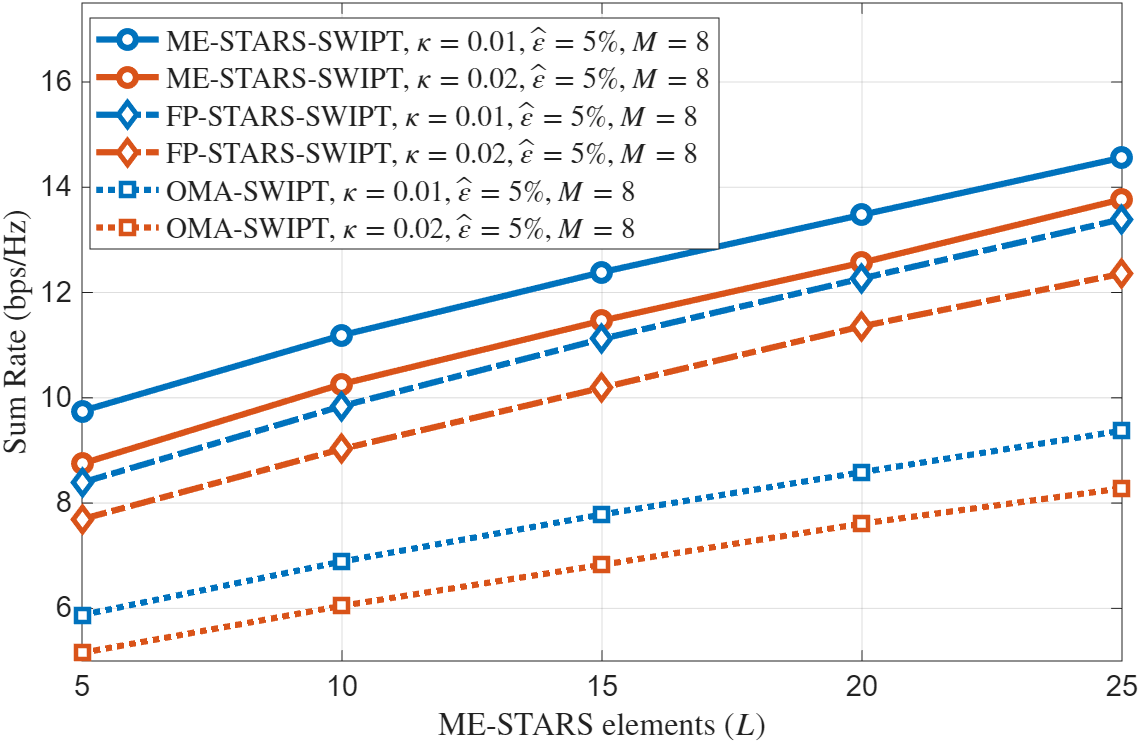}
	\caption{Sum rate versus $L$ under different hardware-impairment levels.}
	\label{f8}
\end{figure} 

\begin{figure}[h]
	\centering
	\includegraphics [width=0.35\textwidth]{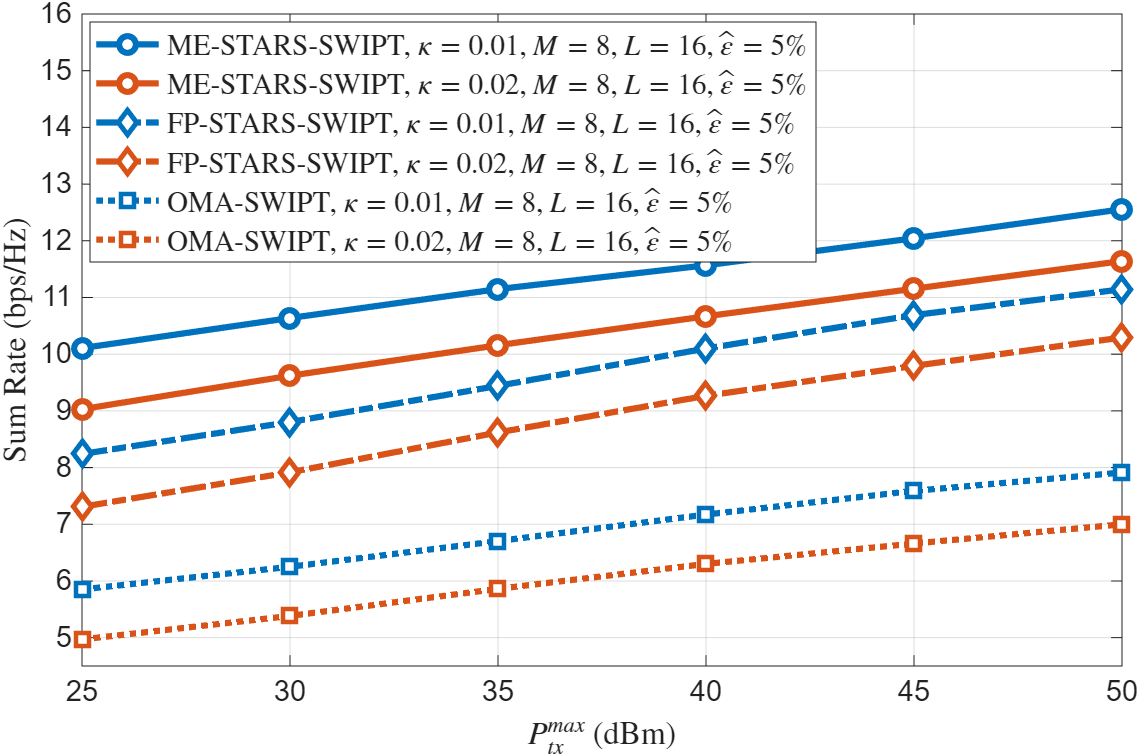}
	\caption{Sum rate versus $P_{\rm tx}^{\max}$ under different hardware-impairment levels.}
	\label{f9}
\end{figure} 

Figs.~\ref{f2}--\ref{f4} illustrate the convergence behavior of the proposed ME-STARS-SWIPT scheme and the fixed-position (FP) STARS-SWIPT benchmark, denoted by FP-STARS-SWIPT, under different CSI uncertainty
levels, hardware-impairment levels, and minimum harvested-power requirements. For the FP-STARS-SWIPT benchmark, the STARS elements are placed at fixed feasible locations and their positions remain unchanged
throughout the optimization process. For all three settings, the sum rate rises quickly in the early iterations and then gradually stabilizes, indicating that the proposed optimization procedure converges within a modest number of iterations. Fig.~\ref{f2} shows a gradual reduction in the achievable sum rate as
$\varepsilon_\nu$ becomes larger, because greater CSI uncertainty makes the estimated cascaded BS--ME-STARS--user channels less reliable, thereby limiting the effectiveness of the jointly designed BS precoders and ME-STARS passive beamforming. Fig.~\ref{f3} further shows that increasing $\kappa$ reduces the achievable sum rate because stronger residual transceiver distortion degrades the effective received-signal quality. Similarly, Fig.~\ref{f4} shows that increasing the EH requirement reduces the achievable sum rate, since a larger fraction of the received RF power must be allocated to energy harvesting to satisfy the higher harvested-power requirement. Despite these adverse effects, the converged sum rate of ME-STARS-SWIPT remains higher than that of FP-STARS-SWIPT for all considered settings, demonstrating the benefit of element-position optimization under CSI uncertainty, residual hardware impairments, and stringent EH requirements.

Fig.~\ref{f5} evaluates how the achievable sum rate varies with the number of STARS elements $L$ for different levels of CSI uncertainty. For both ME-STARS-SWIPT and FP-STARS-SWIPT, the sum rate increases steadily
with $L$, since adding more surface elements offers additional degrees of freedom for configuring the cascaded BS--STARS--user channels. Across all considered values of $L$, ME-STARS-SWIPT outperforms the FP-STARS-SWIPT benchmark in terms of achievable sum rate. This gain stems from the additional spatial freedom provided by element repositioning, which enables the surface to exploit more favorable propagation conditions in addition to optimizing the reflection and transmission coefficients. These results indicate that a larger number of movable elements enhances the spatial beamforming flexibility of the ME-STARS and strengthens the benefit of position optimization in the considered robust RSMA-SWIPT design.

Figs.~\ref{f6} and \ref{f7} examine the achievable sum rate as a function of the maximum BS transmit power under different EH requirements and CSI uncertainty levels, respectively. Increasing $P_{\rm tx}^{\max}$ improves the sum rate of all considered schemes, as the larger transmit-power budget provides greater flexibility for supporting the common and private streams while satisfying the communication and EH requirements. As shown in Fig.~\ref{f6}, reducing the EH requirement improves the achievable sum rate by allowing a smaller share of the received RF power to be allocated to energy harvesting, thereby providing greater flexibility for information decoding. Fig.~\ref{f7} further shows that lower CSI uncertainty leads to higher sum rates, since more reliable knowledge of the cascaded BS--ME-STARS--user channels enables more effective beamforming and STARS configuration. For all considered values of $P_{\rm tx}^{\max}$, ME-STARS-SWIPT consistently outperforms
FP-STARS-SWIPT in terms of achievable sum rate. This gain confirms that optimizing the element positions provides an additional degree of freedom for improving the cascaded channels, allowing the proposed scheme to exploit the available transmit power more effectively under stringent EH requirements and imperfect CSI.

Finally, Figs.~\ref{f8} and \ref{f9} compare the proposed ME-STARS-SWIPT scheme with the FP-STARS-SWIPT and OMA-SWIPT benchmarks under different hardware-impairment levels, considering variations in the number of STARS elements and the maximum BS transmit power, respectively. Here, OMA-SWIPT adopts the same ME-STARS-assisted SWIPT architecture as the proposed scheme, but serves the users according to an orthogonal multiple-access protocol instead of RSMA. As shown in Fig.~\ref{f8}, increasing $L$ improves the achievable sum rate for all schemes because the larger number of STARS elements introduces additional degrees of freedom for configuring the cascaded BS--STARS--user channels. Similarly, Fig.~\ref{f9} shows that larger $P_{\rm tx}^{\max}$ leads to higher sum rates, as the additional transmit-power budget strengthens the useful received signals and provides greater flexibility for meeting the communication and EH requirements. Across the considered values of $L$ and $P_{\rm tx}^{\max}$, ME-STARS-SWIPT achieves the highest sum rate, followed by FP-STARS-SWIPT, whereas OMA-SWIPT yields the lowest performance. The improvement over FP-STARS-SWIPT results from the additional spatial flexibility provided by movable-element positioning, while the performance gap relative to OMA-SWIPT demonstrates the advantage of RSMA in managing multiuser interference through common and private streams.

\section{Conclusion}
We studied robust transmission design for an ME-STARS-enabled RSMA-SWIPT system while accounting for CSI uncertainty and residual transceiver HIs. The proposed optimization framework jointly determines the BS precoders, user-specific common-rate allocation, ME-STARS passive beamforming, PS ratios, and spatial positions of the movable elements so as to maximize the robust sum rate, while accounting for nonlinear energy harvesting and practical system constraints. To address the non-convex formulation, we employ an iterative solution strategy that successively updates the BS precoders, ME-STARS passive beamforming, movable-element positions, and PS ratios through tractable convex reformulations. Numerical results demonstrated the effectiveness and robustness of the proposed framework over the considered benchmark schemes and confirmed its stable convergence under different system configurations.

\bibliographystyle{IEEEtran}
\bibliography{Ref}

@article{wang2022gcwcn,
  title={{GCWCN: 6G-based global coverage wireless communication network architecture}},
  author={Wang, Chao and Zhang, Peiying and Kumar, Neeraj and Liu, Lei and Yang, Tingting},
  journal={IEEE Network},
  volume={37},
  number={3},
  pages={218--223},
  year={2022},
  publisher={IEEE}       
}

@article{nguyen20216g,
  title={{6G Internet of Things: A comprehensive survey}},
  author={Nguyen, Dinh C and Ding, Ming and Pathirana, Pubudu N and Seneviratne, Aruna and Li, Jun and Niyato, Dusit and Dobre, Octavia and Poor, H Vincent},
  journal={IEEE Internet of Things Journal},
  volume={9},
  number={1},
  pages={359--383},
  year={2021},
  publisher={IEEE}
}

@article{zeng2017communications,
  title={{Communications and signals design for wireless power transmission}},
  author={Zeng, Yong and Clerckx, Bruno and Zhang, Rui},
  journal={IEEE Transactions on Communications},
  volume={65},
  number={5},
  pages={2264--2290},
  year={2017},
  publisher={IEEE}
}

@article{park2014joint,
  title={{Joint wireless information and energy transfer in a $ K $-user MIMO interference channel}},
  author={Park, Jaehyun and Clerckx, Bruno},
  journal={IEEE Transactions on Wireless Communications},
  volume={13},
  number={10},
  pages={5781--5796},
  year={2014},
  publisher={IEEE}
}

@article{mao2022rate,
  title={{Rate-splitting multiple access: Fundamentals, survey, and future research trends}},
  author={Mao, Yijie and Dizdar, Onur and Clerckx, Bruno and Schober, Robert and Popovski, Petar and Poor, H Vincent},
  journal={IEEE communications surveys \& tutorials},
  volume={24},
  number={4},
  pages={2073--2126},
  year={2022},
  publisher={IEEE}
}

@article{acosta2020joint,
  title={{Joint power allocation and power splitting for MISO-RSMA cognitive radio systems with SWIPT and information decoder users}},
  author={Acosta, Mario Rodrigo Camana and Moreta, Carla Estefania Garcia and Koo, Insoo},
  journal={IEEE Systems Journal},
  volume={15},
  number={4},
  pages={5289--5300},
  year={2020},
  publisher={IEEE}
}

@article{karim2025finite,
  title={{Finite blocklength analysis for SWIPT-enabled RSMA networks under realistic assumptions}},
  author={Karim, Farjam and Mahmood, Nurul Huda and De Sena, Arthur S and Kumar, Deepak and Latva-aho, Matti},
  journal={IEEE Transactions on Wireless Communications},
  volume={24},
  number={9},
  pages={8014--8024},
  year={2025},
  publisher={IEEE}
}

@article{galappaththige2024sum,
  title={{Sum rate maximization for RSMA-assisted CF mMIMO networks with SWIPT users}},
  author={Galappaththige, Diluka and Tellambura, Chintha},
  journal={IEEE Wireless Communications Letters},
  volume={13},
  number={5},
  pages={1300--1304},
  year={2024},
  publisher={IEEE}
}

@article{asif2026robust,
  title={{Robust Design of Beyond-Diagonal Reconfigurable Intelligent Surface Empowered RSMA-SWIPT System Under Channel Estimation Errors}},
  author={Asif, Muhammad and Ali, Zain and Ihsan, Asim and Ranjha, Ali and Shoujin, Zhu and Ahmed, Manzoor and Li, Xingwang and Chatzinotas, Symeon},
  journal={IEEE Transactions on Wireless Communications},
  year={2026},
  publisher={IEEE}
}

@article{liu2021reconfigurable,
  title={{Reconfigurable intelligent surfaces: Principles and opportunities}},
  author={Liu, Yuanwei and Liu, Xiao and Mu, Xidong and Hou, Tianwei and Xu, Jiaqi and Di Renzo, Marco and Al-Dhahir, Naofal},
  journal={IEEE communications surveys \& tutorials},
  volume={23},
  number={3},
  pages={1546--1577},
  year={2021},
  publisher={IEEE}
}

@article{asif2025noma,
  title={{NOMA-based Ze-RIS empowered backscatter communication with energy-efficient resource management}},
  author={Asif, Muhammad and Bao, Xu and Ihsan, Asim and Khan, Wali Ullah and Li, Xingwang and Chatzinotas, Symeon and Dobre, Octavia A},
  journal={IEEE Transactions on Communications},
  volume={73},
  number={9},
  pages={7193--7209},
  year={2025},
  publisher={IEEE}
}

@article{mu2021simultaneously,
  title={{Simultaneously transmitting and reflecting (STAR) RIS aided wireless communications}},
  author={Mu, Xidong and Liu, Yuanwei and Guo, Li and Lin, Jiaru and Schober, Robert},
  journal={IEEE transactions on wireless communications},
  volume={21},
  number={5},
  pages={3083--3098},
  year={2021},
  publisher={IEEE}
}

@article{asif2026robust123,
  title={{Robust beamforming optimization for STAR-RIS empowered multi-user RSMA under hardware imperfections and channel uncertainty}},
  author={Asif, Muhammad and Ihsan, Asim and Shoujin, Zhu and Ranjha, Ali and Li, Xingwang and Rabie, Khaled M and Chatzinotas, Symeon},
  journal={IEEE Transactions on Communications},
  year={2026},
  publisher={IEEE}
}

@article{hashempour2024secure,
  title={{Secure SWIPT in the multiuser STAR-RIS aided MISO rate splitting downlink}},
  author={Hashempour, Hamid Reza and Bastami, Hamed and Moradikia, Majid and Zekavat, Seyed A and Behroozi, Hamid and Berardinelli, Gilberto and Swindlehurst, A Lee},
  journal={IEEE Transactions on Vehicular Technology},
  volume={73},
  number={9},
  pages={13466--13481},
  year={2024},
  publisher={IEEE}
}

@article{amiri2025resource,
  title={{Resource allocation in STAR-RIS-aided SWIPT with RSMA via meta-learning}},
  author={Amiri, Mojtaba and Vaezpour, Elaheh and Javadi, Sepideh and Mili, Mohammad Robat and Bennis, Mehdi and Jorswieck, Eduard Axel},
  journal={IEEE Open Journal of the Communications Society},
  volume={6},
  pages={3806--3815},
  year={2025},
  publisher={IEEE}
}

@article{asif2024leveraging,
  title={{Leveraging RIS in consumer-centric 6G networks: Efficient resource allocation in RSMA-based SWIPT systems under hardware impairments}},
  author={Asif, Muhammad and Bao, Xu and Ranjha, Ali and Ahmed, Manzoor and Khan, Wali Ullah and Rani, Shalli and Li, Xingwang},
  journal={IEEE Transactions on Consumer Electronics},
  volume={71},
  number={2},
  pages={4235--4247},
  year={2024},
  publisher={IEEE}
}

@article{zhu2023modeling,
  title={{Modeling and performance analysis for movable antenna enabled wireless communications}},
  author={Zhu, Lipeng and Ma, Wenyan and Zhang, Rui},
  journal={IEEE Transactions on Wireless Communications},
  volume={23},
  number={6},
  pages={6234--6250},
  year={2023},
  publisher={IEEE}
}

@article{hokmabadi2026joint,
  title={{Joint Beamforming and Position Optimization for Movable-Antenna and Movable-Element RIS--Aided Full-Duplex 6G MISO Systems}},
  author={Hokmabadi, Ayda Nodel and Assi, Chadi},
  journal={IEEE Transactions on Communications},
  year={2026},
  publisher={IEEE}
}

@article{zhou2025movable,
  title={{Movable-Element RIS: Joint Element Positioning and Beamforming Optimization}},
  author={Zhou, Di and Mei, Weidong and Bai, Zhiquan and Li, Na and Quek, Tony QS},
  journal={IEEE Wireless Communications Letters},
  volume={15},
  pages={915--919},
  year={2025},
  publisher={IEEE}
}

@article{hu2024intelligent,
  title={{Intelligent reflecting surface-aided wireless communication with movable elements}},
  author={Hu, Guojie and Wu, Qingqing and Xu, Donghui and Xu, Kui and Si, Jiangbo and Cai, Yunlong and Al-Dhahir, Naofal},
  journal={IEEE Wireless Communications Letters},
  volume={13},
  number={4},
  pages={1173--1177},
  year={2024},
  publisher={IEEE}
}

@article{zhao2026movable,
  title={{Movable-element RIS-aided wireless communications: An element-wise position optimization approach}},
  author={Zhao, Jingjing and Huang, Qingyi and Cai, Kaiquan and Zhou, Quan and Mu, Xidong and Liu, Yuanwei},
  journal={IEEE Communications Letters},
  year={2026},
  publisher={IEEE}
}

@article{zhu2025movable,
  title={{Movable-element STARS-assisted near-field wideband communications}},
  author={Zhu, Guangyu and Mu, Xidong and Guo, Li and Huang, Ao and Xu, Shibiao},
  journal={IEEE Internet of Things Journal},
  volume={12},
  number={16},
  pages={33130--33143},
  year={2025},
  publisher={IEEE}
}

@article{zhao2025movable,
  title={{Movable-element STARS-aided secure communications}},
  author={Zhao, Jingjing and Xu, Qian and Cai, Kaiquan and Zhu, Yanbo and Mu, Xidong and Liu, Yuanwei},
  journal={IEEE Transactions on Vehicular Technology},
  year={2025},
  publisher={IEEE}
}

@article{asif2026exploiting,
  title={{Exploiting Movable-Element STARS for Rate Splitting Multiple Access}},
  author={Asif, Muhammad and Ihsan, Asim and Muhammad, Irfan and Shaikh, Mohd Hamza Naim and Mirza, Muhammad Ayzed and Shoujin, Zhu and Chatzinotas, Symeon},
  journal={arXiv preprint arXiv:2608.16866},
  year={2026}
}

@article{zhao2026exploiting,
  title={{Exploiting movable-element STARS for wireless communications}},
  author={Zhao, Jingjing and Zhou, Quan and Mu, Xidong and Cai, Kaiquan and Zhu, Yanbo and Liu, Yuanwei},
  journal={IEEE Transactions on Wireless Communications},
  year={2026},
  publisher={IEEE}
}

@article{liu2026joint,
  title={{Joint Beamforming and Position Optimization for FIRES-NOMA Assisted Wireless Communication Systems}},
  author={Liu, Yu and Luo, Qu and Chen, Gaojie and Xiao, Pei and Elzanaty, Ahmed and Khalily, Mohsen and Tafazolli, Rahim},
  journal={IEEE Transactions on Communications},
  year={2026},
  publisher={IEEE}
}

@article{shen2020beamforming,
  title={{Beamforming optimization for IRS-aided communications with transceiver hardware impairments}},
  author={Shen, Hong and Xu, Wei and Gong, Shulei and Zhao, Chunming and Ng, Derrick Wing Kwan},
  journal={IEEE Transactions on Communications},
  volume={69},
  number={2},
  pages={1214--1227},
  year={2020},
  publisher={IEEE}
}

@article{zhang2023robust,
  title={{Robust beamforming design for RIS-aided NOMA secure networks with transceiver hardware impairments}},
  author={Zhang, Qian and Liu, Ju and Gao, Zhichao and Li, Ziyu and Peng, Zhiying and Dong, Zheng and Xu, Hongji},
  journal={IEEE Transactions on Communications},
  volume={71},
  number={6},
  pages={3637--3649},
  year={2023},
  publisher={IEEE}
}

@article{yao2023robust,
  title={Robust beamforming design for RIS-aided cell-free systems with CSI uncertainties and capacity-limited backhaul},
  author={Yao, Jiacheng and Xu, Jindan and Xu, Wei and Ng, Derrick Wing Kwan and Yuen, Chau and You, Xiaohu},
  journal={IEEE Transactions on Communications},
  volume={71},
  number={8},
  pages={4636--4649},
  year={2023},
  publisher={IEEE}
}

@article{boshkovska2015practical,
  title={{Practical non-linear energy harvesting model and resource allocation for SWIPT systems}},                      
  author={Boshkovska, Elena and Ng, Derrick Wing Kwan and Zlatanov, Nikola and Schober, Robert},
  journal={IEEE Communications Letters},
  volume={19},
  number={12},
  pages={2082--2085},
  year={2015},
  publisher={IEEE}
}

@article{xiong2017rate,
  title={Rate-energy region of SWIPT for MIMO broadcasting under nonlinear energy harvesting model},
  author={Xiong, Ke and Wang, Beibei and Liu, KJ Ray},
  journal={IEEE Transactions on Wireless Communications},
  volume={16},
  number={8},
  pages={5147--5161},
  year={2017},
  publisher={IEEE}
}

@article{ni2021resource,
  title={{Resource allocation for multi-cell IRS-aided NOMA networks}},
  author={Ni, Wanli and Liu, Xiao and Liu, Yuanwei and Tian, Hui and Chen, Yue},
  journal={IEEE Transactions on Wireless Communications},
  volume={20},
  number={7},
  pages={4253--4268},
  year={2021},
  publisher={IEEE}
}

@article{luo2010semidefinite,
  title={{Semidefinite relaxation of quadratic optimization problems}},
  author={Luo, Zhi-Quan and Ma, Wing-Kin and So, Anthony Man-Cho and Ye, Yinyu and Zhang, Shuzhong},
  journal={IEEE Signal Processing Magazine},
  volume={27},
  number={3},
  pages={20--34},
  year={2010},
  publisher={IEEE}
}

@misc{grant2008cvx,
  title={{CVX: Matlab software for disciplined convex programming}},
  author={Grant, Michael and Boyd, Stephen and Ye, Yinyu},
  year={2008},
  publisher={Stanford University, Stanford, CA, USA}
}

@article{geng2025joint,
  title={Joint beamforming and antenna position optimization for IRS-aided multi-user movable antenna systems},
  author={Geng, Yue and Cheng, Tee Hiang and Zhong, Kai and Teh, Kah Chan and Wu, Qingqing},
  journal={IEEE Transactions on Wireless Communications},
  year={2025},
  publisher={IEEE}
}
\vskip -2\baselineskip plus -2fil

\end{document}